\pdfoutput=1
\documentclass[aps,prd,onecolumn,groupedaddress,floatfix,longbibliography,superscriptaddress,notitlepage]{revtex4-2}

\usepackage{amsfonts}

\usepackage{amsmath}
\usepackage{amssymb}
\usepackage{amsfonts}
\usepackage{bbold}
\usepackage{bm}
\usepackage{cancel}
\usepackage{times,float}
\usepackage{graphicx}
\usepackage[dvipsnames,svgnames]{xcolor}
\usepackage{hyperref}
\hypersetup{colorlinks=true, linkcolor=Blue, citecolor=Blue,urlcolor=Blue}
\usepackage{multirow}
\usepackage{ulem}
\usepackage{color,epsfig}
\usepackage{bm}
\usepackage{slashed}
\usepackage{hyperref}
\usepackage{feynmf}
\usepackage{array}
\usepackage{physics}
\usepackage{subcaption}
\allowdisplaybreaks
\usepackage{comment}
\usepackage[none]{hyphenat}

\begin{document}

\title{Probing Lorentz symmetry violation via the Casimir effect in rectangular cavities}

\author{M. B. Cruz}
\email{messiasdebritocruz@servidor.uepb.edu.br}
\affiliation{Universidade Estadual da Paraíba (UEPB), Centro de Ciências Exatas e Sociais Aplicadas (CCEA), R. Alfredo Lustosa Cabral, s/n, Salgadinho, Patos - PB, 58706-550 - Brazil.}

\author{E. R. Bezerra de Mello}
\email{emello@fisica.ufpb.br}
\affiliation{Departamento de Física, Universidade Federal da Paraíba,
Caixa Postal 5008, Jo\~ao Pessoa, Paraíba, Brazil}

\author{A. Mart\'{i}n-Ruiz}
\email{alberto.martin@nucleares.unam.mx}
\affiliation{Instituto de Ciencias Nucleares, Universidad Nacional Aut\'{o}noma de M\'{e}xico, 04510 Ciudad de M\'{e}xico, M\'{e}xico}

\begin{abstract}
\noindent
We investigate the Casimir effect as a probe of Lorentz symmetry violation for a real scalar field confined to a rectangular waveguide with Dirichlet boundary conditions. The field dynamics is governed by a Lorentz-violating extension of the Klein-Gordon theory involving a fixed background four-vector $u_\mu$. Focusing on four representative configurations in which the background is aligned with the temporal direction or with one of the spatial axes of the cavity, we derive the modified mode spectra and the corresponding vacuum energies. We show that these configurations induce anisotropic modifications of the dispersion relations that depend explicitly on the orientation of the background vector relative to the cavity geometry, while still preserving mode separability. The resulting Casimir energy acquires characteristic direction-dependent corrections that encode the breaking of Lorentz symmetry, without altering the universal functional structure of the spectral kernel. Our analysis provides a controlled and transparent framework for isolating Lorentz-violating effects in confined geometries and highlights Casimir systems as sensitive probes of anisotropic physics and fundamental spacetime symmetries.
\end{abstract}


\maketitle

\section{Introduction} \label{intro}

The Casimir effect, originally predicted by H.~B.~G.~Casimir in 1948 \cite{Casimir:1948dh,Milton:2001yy}, is one of the clearest manifestations of quantum vacuum fluctuations in the presence of boundaries. Early measurements provided only limited accuracy \cite{Sparnaay:1958wg}, but subsequent high-precision experiments \cite{Lamoreaux:1996wh,Mohideen:1998iz,Decca:2003zk,Bressi:2002fr} firmly established the effect and turned it into a valuable laboratory for precision tests of quantum field theory (QFT) and of fluctuation-induced forces \cite{Bordag:2009zz,Milton:2001yy,Plunien:1986ca,Klimchitskaya:2009zz}. In its simplest form, the Casimir effect arises because boundary conditions modify the normal-mode spectrum of a confined quantum field, leading to observable stresses and forces between material surfaces \cite{Casimir:1948dh,Bordag:2009zz}. From a theoretical perspective, the phenomenon extends far beyond the original electromagnetic parallel-plate setup: it has been analyzed for different quantum fields, geometries, and boundary conditions, and it also admits natural generalizations to nontrivial backgrounds, including curved spacetimes \cite{Bordag:2009zz,Birrell:1982ix}.

Quantum field theory (QFT) relies fundamentally on spacetime symmetries, among which Lorentz invariance and CPT symmetry play a central role in ensuring the consistency of relativistic field dynamics and particle interactions. Nevertheless, several candidate theories beyond the Standard Model predict that these symmetries may be violated at sufficiently high energies. For example, spontaneous Lorentz symmetry breaking arises naturally in string-inspired scenarios \cite{Kostelecky:1988zi}, while the general framework of the Standard-Model Extension (SME) provides a systematic effective-field-theory description of Lorentz- and CPT-violating operators compatible with known interactions \cite{Colladay:1996iz,Colladay:1998fq,Kostelecky:2003fs,Kostelecky:2009zp}. In addition, modified gravity models and quantum-gravity–motivated frameworks predict preferred spacetime structures that lead to anisotropic propagation and modified dispersion relations for quantum fields \cite{Jacobson:2000xp,Mattingly:2005re,Liberati:2013xla}. These developments establish Lorentz symmetry violation as a theoretically well-motivated possibility and provide a consistent framework for investigating its phenomenological consequences.

In this context, the Casimir effect has emerged as a particularly sensitive probe of physics beyond the Standard Model. Since the Casimir force originates directly from the vacuum mode spectrum of a confined quantum field, it is inherently sensitive to modifications of dispersion relations, boundary conditions, and spacetime structure. For this reason, Casimir energies have been investigated in a wide range of extensions of conventional QFT, including theories with nontrivial topology \cite{Martin_2016, BezerradeMello:2011nv}, modified gravitational backgrounds \cite{Sorge_2005, Nazari_EPJC_2015, Quach:2015qwa, deMello:2011mw, BezerradeMello:2017nyo, Muniz_2025}, discrete or polymer quantization schemes \cite{Escobar:2020zat, Ishikawa:2020icy}, and other generalized field-theory frameworks \cite{Bezerra:2014pza, MoralesUlion:2015tve, Ferrari:2010dj, deMello:2024lmi, ESCOBAR2024169570, Linares:2007yz, Linares:2010uy, MUNIZ2024101673}. In such settings, the Casimir effect provides direct access to the spectral properties of vacuum fluctuations and offers a robust theoretical tool to test deviations from standard relativistic dynamics.

In addition to its role as a probe of fundamental physics, the Casimir effect has important practical implications. At submicron scales, fluctuation-induced forces become comparable to mechanical forces and play a crucial role in the operation of micro- and nanoelectromechanical systems (MEMS and NEMS), where they can significantly influence stability, adhesion, and device performance \cite{Capasso:2007zz,Bordag:2009zz}. Consequently, modifications of the vacuum spectrum induced by spacetime anisotropies or other fundamental effects may, in principle, provide new mechanisms for controlling Casimir forces, opening the possibility of engineered vacuum interactions and novel technological applications \cite{Intravaia:2015nea,Woods:2016vwu}.

Within the SME, Lorentz violation is described by fixed background tensor coefficients coupled to the photon, scalar, fermion, and gravitational sectors \cite{Colladay:1996iz,Colladay:1998fq,Kostelecky:2003fs,Kostelecky:2009zp}. These coefficients modify the kinetic structure of the theory and lead to anisotropic propagation and altered dispersion relations. In CPT-even ether-type models, such modifications directly affect the vacuum mode spectrum and the associated zero-point energy. Because the Casimir effect depends explicitly on the spectrum of confined modes, it provides a natural observable to probe these deviations. Consequently, Casimir configurations, particularly the parallel-plate geometry, have been extensively analyzed in Lorentz-violating scalar \cite{Cruz:2017kfo, Cruz:2018bqt, Escobar:2020pes, ESCOBAR2020135567, Martin-Ruiz:2020lfo, Escobar-Ruiz:2021dxi, lz5l-f6kd}, fermionic \cite{Frank:2006ww, Cruz:2018thz, Rohim:2023tmy}, and electromagnetic field theories \cite{Kharlanov:2009pv, Martin-Ruiz:2016ijc, Martin-Ruiz:2016lyy}, demonstrating the sensitivity of vacuum energies to small departures from Lorentz invariance.

Motivated by these considerations, in this work we investigate the Casimir effect for a real massive scalar field confined within a rectangular waveguide of cross section $L_x\times L_y$, corresponding to a system bounded by four parallel plates and infinite along the longitudinal direction. We consider a CPT-even aether-type Lorentz-violating extension of the Klein-Gordon theory within the effective-field-theory framework \cite{Colladay:1996iz,Colladay:1998fq,Kostelecky:2003fs}, in which Lorentz violation is encoded through a fixed background four-vector that induces anisotropic modifications of the dispersion relation. For aligned configurations of the background, the spectral problem remains separable and the Lorentz-violating effects appear as direction-dependent rescalings of the propagation and confinement scales. The Casimir energy is defined through a renormalized sum-minus-integral prescription and evaluated by means of a twofold application of the Abel-Plana summation formula, which isolates the finite interaction energy and allows the discrete mode sums to be expressed in terms of exponentially convergent series of modified Bessel functions. This yields an exact closed representation of the Casimir energy density valid for arbitrary mass, cavity dimensions, and Lorentz-violating parameters, from which the dependence of the vacuum energy on the anisotropic coefficients and the geometric scales can be analyzed in both the small- and large-mass regimes.

This article is organized as follows. In Section \ref{QFT_model}, we introduce the theoretical framework, presenting the massive scalar field model within the aether-type LV background. In Section \ref{casimir_LV}, we determine the spectrum of normalizable modes and construct the corresponding Casimir energy for the four-plate geometry, considering both Dirichlet boundary conditions along the confined directions. The divergences inherent to the mode summation are handled using the Abel-Plana summation formula, leading to finite and physically meaningful results. Finally, in Section \ref{conclusions}, we discuss our findings and their physical implications. Throughout the paper, we adopt the metric signature $(+,-,-,-)$.

\section{Scalar field theory with Lorentz symmetry violation} \label{QFT_model}

In this section, we introduce the theoretical framework that underlies our entire analysis. We consider a real massive scalar quantum field theory modified by the inclusion of an aether-type term, which explicitly breaks Lorentz symmetry through its coupling to a fixed background four-vector. Such extensions are characteristic of effective field theories with LV backgrounds and provide a systematic framework to investigate possible deviations from exact Lorentz invariance \cite{Colladay:1998fq,Kostelecky:2003fs}.

The dynamics of the scalar field are governed by the following modified Klein-Gordon Lagrangian density:
\begin{align}
\mathcal{L} = \frac{1}{2} \left[ \frac{1}{c^{2}}(\partial _{t} \phi) ^{2} - (\nabla \phi) ^{2} + \Lambda \,(u ^{\mu} \partial _{\mu} \phi ) ^{2} + \left( \frac{mc}{\hbar} \right) ^{2} \phi ^{2} \right] .  \label{Lagrangian}
\end{align}
Here, $\phi$ denotes a real scalar field of mass $m$. The LV contribution is encoded in the derivative coupling to a constant background four-vector $u^{\mu}$, while the dimensionless parameter $\Lambda$ controls the strength of the LV and is assumed to be much smaller than unity throughout this work. The presence of the fixed vector $u^{\mu}$ explicitly breaks invariance under Lorentz transformations, whereas translational invariance is preserved provided that $u^{\mu}$ is spacetime independent. Background structures of this type arise naturally in effective field-theory descriptions of LV physics and have been widely employed in phenomenological studies \cite{Colladay:1998fq,Kostelecky:2003fs}.

The equation of motion follows from the Euler-Lagrange equations. By performing the variation explicitly and taking into account Eq. \eqref{Lagrangian}, we obtain a modified Klein-Gordon equation of the form:
\begin{align} \label{Modified_KG}
\Big[ \Box + \Lambda \, (u \cdot \partial ) ^{2} + \left( \frac{mc}{\hbar} \right) ^{2} \Big] \phi = 0 , \qquad (u \cdot \partial ) \equiv u ^{\mu} \partial _{\mu} , 
\end{align}
where $\Box = \frac{1}{c ^{2} } \partial_{t} ^{2} - \nabla ^{2}$ denotes the d'Alembert operator. The energy--momentum tensor, $T_{\mu\nu}$, plays an important role in characterizing the flow of energy and momentum in the theory. Considering a Lagrangian density, the canonical energy-momentum tensor is defined as:
\begin{align}
T ^{\mu \nu} = \frac{\partial \mathcal{L} }{\partial(\partial _{\mu} \phi)} \, \partial ^{\nu} \phi - \eta ^{\mu\nu} \mathcal{L} . \label{EM_tensor_def}
\end{align}
For the scalar field modified by LV considered here, this expression becomes:
\begin{align}
T ^{\mu\nu} = (\partial ^{\mu} \phi ) ( \partial ^{\nu} \phi ) +  \Lambda \, u ^{\mu} ( \partial ^{\nu} \phi ) ( u \cdot \partial \phi ) - \eta ^{\mu\nu} \mathcal{L} . \label{EM_tensor}
\end{align}

It is important to emphasize that the violation of Lorentz symmetry does not, by itself, preclude the conservation of energy and momentum. As long as the background vector $u ^{\mu}$ is constant, translational invariance is preserved and the energy-momentum tensor remains conserved, $\partial _{\mu} T ^{\mu \nu} = 0$. However, the presence of the Lorentz-violating term modifies the tensor structure and, in particular, renders the energy-momentum tensor generically
non-symmetric. Its antisymmetric part is given by
\begin{align}
T ^{\mu\nu} - T ^{\nu\mu} =  \Lambda \left[ u ^{\mu} ( \partial ^{\nu} \phi )  - u ^{\nu} ( \partial ^{\mu} \phi ) \right] (u \cdot \partial \phi ) .  \label{EM_antisymmetric}
\end{align}
This non-symmetric feature of the energy-momentum tensor reflects the explicit LV induced by the background vector. Such a property may lead to observable consequences in physical systems where boundary conditions or external constraints select preferred directions. In particular, high-precision phenomena sensitive to vacuum fluctuations, such as the Casimir effect \cite{Frank:2006ww, Kharlanov:2009pv, Cruz:2018thz}, provide a natural arena to investigate the interplay between geometry and Lorentz symmetry violation. Related signatures have also been discussed in the context of accelerator experiments \cite{Kostelecky:2008ts, Mattingly:2005re}, astrophysical observations \cite{Jacobson:2006gn, Liberati:2013xla}, black-hole physics \cite{Blas:2011ni, Barausse:2011mp}, and cosmology \cite{Kostelecky:2010ze, Mewes:2012ph}.

\section{Impact of Lorentz violation on the Casimir effect} \label{casimir_LV}

In this section, we investigate the influence of LV on the Casimir effect. As discussed previously, the anisotropy introduced by this violation is controlled by the parameter $\Lambda$, appearing in Eq. \eqref{Modified_KG}, which modifies the dispersion relation of the field. Consequently, deviations in the Casimir energy and, therefore, in the Casimir pressure are naturally expected when compared to the isotropic scenario, namely, in the absence of Lorentz violation.

\begin{figure}[h!]
    \centering
    \includegraphics[width=0.5\linewidth]{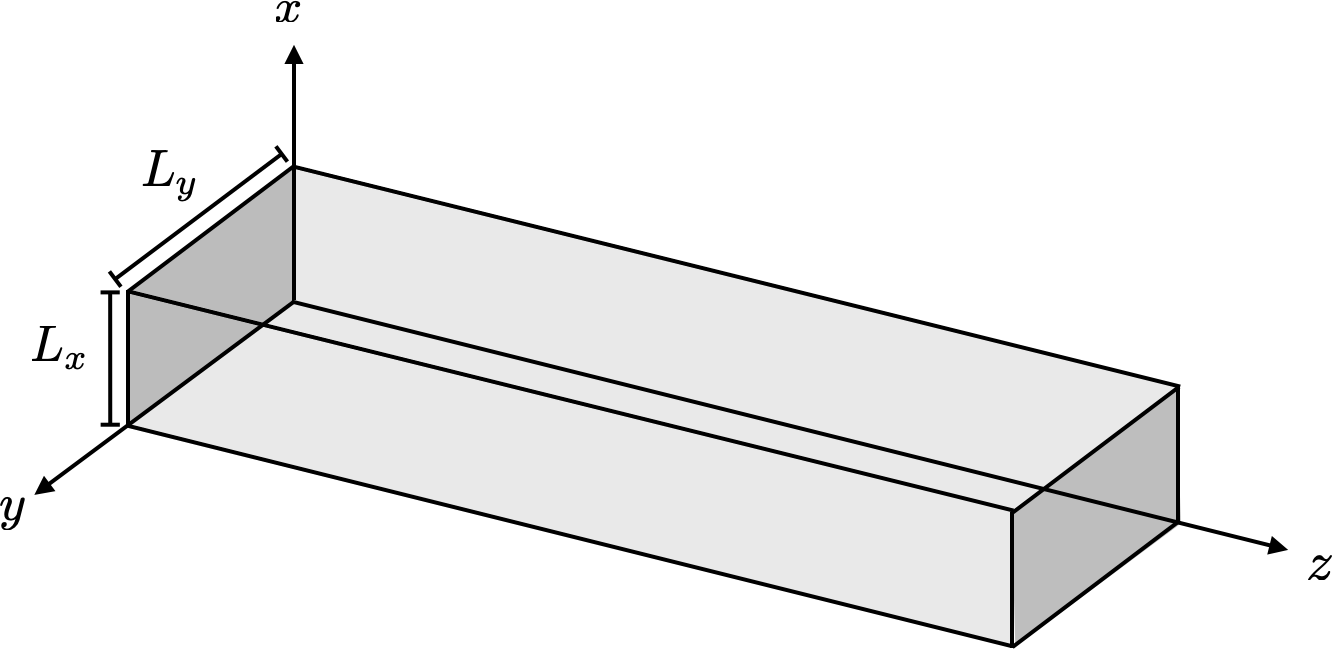}
    \caption{Rectangular cavity formed by parallel plates located at $x=0, \ L_x$ and $y=0, \ L_y$, confining a massive scalar field $\phi(x,y,z)$ with Dirichlet boundary conditions. The field is free along the $z$-direction, which is taken to be unbounded.}
    \label{fig_rec_cav}
\end{figure}

We consider a massive scalar field confined between parallel plates forming a rectangular cavity, as illustrated in Fig. \ref{fig_rec_cav}, and subject to Dirichlet boundary conditions on the cavity walls. This choice allows us to analyze the Casimir effect in a setup close to the well-known two-plate configuration, which has been extensively studied in the literature and enables meaningful comparisons.

In the following subsections, we assume that the four-vector $u^{\mu}$ appearing in Eq. \eqref{Modified_KG} can be either timelike or spacelike. We first obtain the normalized solutions of the modified Klein-Gordon equation that satisfy the Dirichlet boundary conditions on the cavity walls. Subsequently, we construct the corresponding Hamiltonian operator and evaluate its vacuum energy. After implementing the regularization and renormalization procedures, we determine the Casimir energy and analyze the influence of both the cavity geometry and the LV parameter.

\subsection{Geometry, boundary conditions, and mode basis} \label{subsec:geometry_bc_modes}

Considering a rectangular waveguide (i.e., a cavity of infinite extent along the $z$ direction), we define:
\begin{align}
0<x<L_{x} , \qquad 0<y<L_{y} , \qquad z \in \mathbb{R} . \label{region_waveguide}
\end{align}
On the lateral walls, we impose Dirichlet boundary conditions:
\begin{align}
\phi(0,y,z,t)=\phi(L_x,y,z,t)=\phi(x,0,z,t)=\phi(x,L_y,z,t)=0 ,
\label{Dirichlet_BCs}
\end{align}
with the field free to propagate along the longitudinal direction 
$z$.

The transverse eigenfunctions satisfying the modified Klein-Gordon equation, Eq. \eqref{Modified_KG}, under Dirichlet boundary conditions, Eq. \eqref{Dirichlet_BCs}, are given by:
\begin{align}
u _{\mathbf{n}}(x,y) = \sqrt{\frac{4}{L _{x} L _{y} }} \, \sin \left( \frac{n _{x} \pi x}{L _{x} } \right) \sin \left(\frac{n _{y} \pi y}{L _{y} } \right) ,  \qquad \mathbf{n} = (n _{x} , n _{y} ) \in \mathbb{N} ^{2}, \label{transverse_modes_dirichlet}
\end{align}
which satisfy the orthonormality relation
\begin{align}
\int _{0} ^{L _{x} } dx \int _{0} ^{L _{y} }  dy \; u _{\mathbf{n}}(x,y) u _{\mathbf{m}} (x,y) = \delta _{\mathbf{n} \mathbf{m}} .  \label{orthonormality_transverse}
\end{align}

Along the unbounded longitudinal $z$ direction, the field is described by plane-wave modes of the form:
\begin{align}
\varphi _{k} (z) = \frac{e ^{ikz}}{\sqrt{2 \pi}}, \qquad k \in \mathbb{R}. \label{longitudinal_modes}
\end{align}
Consequently, by expanding the real massive scalar field in terms of normal modes, with mode coefficients $\phi_{\mathbf{n}}(k,t)$, we obtain:
\begin{align}
\phi(x,y,z,t) = \sum _{\mathbf{n}} \int _{- \infty} ^{\infty} \frac{dk}{\sqrt{2\pi}} \; \phi _{\mathbf{n}}(k,t) \, u _{\mathbf{n}}(x,y)\, e ^{ikz} . \label{field_expansion_general}
\end{align}

\subsection{Separable Lorentz-violating configurations and dispersion relations}
\label{subsec:separable_cases}

The LV operator $(u \cdot \partial)^2$ in Eq. \eqref{Modified_KG} generally mixes transverse modes when the four-vector $u^{\mu}$ has components orthogonal to the cavity axes. However, we focus on special cases in which the spectrum remains separable and each mode can still be labeled by $(\mathbf{n},k)$. In these cases, the field operator can be quantized in close analogy with the Lorentz-invariant theory, with the LV being incorporated exclusively through a modified dispersion relation:
\begin{align}
\omega _{\mathbf{n}}(k) \; \longrightarrow \; \omega _{\mathbf{n}} ^{(\lambda)} (k) ,
\end{align}
where $(\lambda)$ labels the chosen configuration of $u ^{\mu}$. For future reference, we define
\begin{align}
\xi _{\mathbf{n}} ^{2} = \left( \frac{n _{x} \pi}{L _{x}} \right) ^{2} + \left( \frac{n _{y} \pi}{L _{y}} \right) ^{2} + \left( \frac{mc}{\hbar} \right) ^{2} . \label{mu_n_def}
\end{align}

\begin{itemize}
    \item[\textbf{Case I:}] Timelike background $u^{\mu} = (u _{0},0,0,0)$. In this case $(u\cdot\partial)=(u_0/c)\partial_t$ and the equation of motion implies a rescaling of the temporal kinetic term. The dispersion relation takes the form
    \begin{align}
    \omega _{\mathbf{n}} ^{(\text{I})} (k) = \frac{c}{\sqrt{1+ \Lambda u _{0} ^{2}}} \, \sqrt{k ^{2} + \xi _{\mathbf{n}} ^{2} } . \label{disp_caseI}
    \end{align}

    \item[\textbf{Case II:}] Spacelike background along $x$, $u^\mu=(0,u _{1},0,0)$. Here $(u\cdot\partial)=u_1\partial_x$ and the Lorentz-violating term rescales the $x$-confinement contribution, yielding
    \begin{align}
    \omega _{\mathbf{n}} ^{(\text{II})}(k) = c \, \sqrt{ k ^{2} + (1 -  \Lambda u _{1} ^{2} ) \left( \frac{n _{x} \pi}{L _{x}} \right) ^{2} + \left( \frac{n _{y} \pi}{L _{y}} \right) ^{2} + \left( \frac{mc}{\hbar} \right) ^{2} } . \label{disp_caseII}
    \end{align}

    \item[\textbf{Case III:}] Spacelike background along $y$, $u^\mu=(0,0,u_2,0)$. Analogously to Case II, the dispersion relation reads
    \begin{align}
    \omega _{\mathbf{n}} ^{(\text{III})}(k) = c \, \sqrt{ k ^{2} + \left( \frac{n _{x}\pi}{L _{x}} \right) ^{2} + ( 1 - \Lambda u _{2} ^{2} ) \left( \frac{n _{y} \pi}{L _{y}} \right) ^{2} + \left( \frac{mc}{\hbar} \right) ^{2} } .  \label{disp_caseIII}
    \end{align}

    \item[\textbf{Case IV:}] Spacelike background along $z$, $u^\mu=(0,0,0,u _{3})$. In this case, $(u \cdot \partial) = u_{3} \partial_{z}$, and the longitudinal momentum is anisotropically rescaled as
    \begin{align}
    \omega_{\mathbf n}^{(\text{IV})}(k) = c \, \sqrt{ (1 - \Lambda u _{3} ^{2} ) k ^{2} + \xi _{\mathbf{n}} ^{2} } .  \label{disp_caseIV}
    \end{align}
\end{itemize}

In all four configurations above, LV manifests itself through explicit anisotropic renormalizations of either the temporal kinetic term (Case I), the transverse confinement scales (Cases II and III), or the longitudinal propagation (Case IV). Stability requires the corresponding kinetic coefficients to remain positive, namely $1 + \Lambda u_0^2 > 0$ and $1 - \Lambda u_i^2 > 0$ for the relevant spacelike components. For later convenience, we introduce a generic square-root form spectrum:
\begin{align}
\omega_{\mathbf n} ^{(\lambda)} (k) = c \, A ^{(\lambda)} \, \sqrt{ B ^{(\lambda)} \,  k ^{2} + C ^{(\lambda)} \, \left( \frac{n _{x} \pi}{L _{x}} \right) ^{2} + D ^{(\lambda)} \, \left( \frac{n _{y} \pi}{L _{y}} \right) ^{2} + \left( \frac{mc}{\hbar} \right) ^{2} } .  \label{disp_general_case}
\end{align}
Here, the dimensionless coefficients $A^{(\lambda)}$, $B^{(\lambda)}$, $C^{(\lambda)}$, and $D^{(\lambda)}$ encode the anisotropic effects induced by the Lorentz-violating background. These coefficients are read off directly from Eqs. \eqref{disp_caseI}-\eqref{disp_caseIV}. 

Specifically, Case I corresponds to a global rescaling of the frequency, $A^{(\text{I})} = (1 + \Lambda u_0^{2})^{-1/2}$, with all other coefficients equal to unity. Cases II and III modify the transverse confinement scales through $C^{(\text{II})} = 1 - \Lambda u_1^{2}$ or $D^{(\text{III})} = 1 - \Lambda u_2^{2}$, respectively,
while Case IV affects the longitudinal propagation via $B^{(\text{IV})} = 1 - \Lambda u_3^{2}$. In this way, all separable LV configurations can be treated on equal footing within a unified spectral representation, which provides the starting point for the Abel-Plana analysis of the Casimir energy developed in the following sections.

For a generic orientation of the background vector $u^\mu$, the operator $(u \cdot \partial)^2$ generates mixed derivative terms that couple different transverse modes. In this situation the field equation is no longer separable, and the spectral problem becomes genuinely matrix-valued in the space of cavity eigenfunctions. Such mode mixing is expected to lift degeneracies and to produce contributions to the Casimir energy that cannot be absorbed into simple geometric rescalings of the cavity dimensions. Although this generic scenario lies beyond the scope of the present work, the aligned configurations studied here provide a well-defined and analytically tractable framework in which the impact of LV on vacuum fluctuations can be isolated and characterized unambiguously.

\subsection{Field quantization and vacuum energy} \label{subsec:quantization_vacuum}

Now, for each separable configuration $\lambda = \mathrm{I, II, III, IV}$, the field operator admits the standard normal-mode expansion
\begin{align}
    \phi(x,y,z,t) = \sum _{\mathbf{n}} \int _{- \infty} ^{\infty} dk \; \sqrt{\frac{\hbar c ^{2}}{4 \pi \,\omega _{\mathbf{n}} ^{(\lambda)} (k)}} \; u _{\mathbf{n}} (x,y) \, \Big[ \hat{a} _{\mathbf{n}} (k) \, e ^{i(kz-\omega _{\mathbf{n}} ^{(\lambda)} (k) t )} + \hat{a} _{\mathbf{n}} ^{\dagger} (k) \, e ^{-i(kz- \omega _{\mathbf{n}} ^{(\lambda)} (k) t ) }  \Big] ,  \label{field_operator_separable}
\end{align}
where $\hat{a} _{\mathbf{n}}(k)$ and $\hat{a} _{\mathbf{n}} ^{\dagger}(k)$ satisfy the usual bosonic algebra:
\begin{align}
    [ \hat{a} _{\mathbf{n}} (k) , \hat{a} _{\mathbf{n}} ^{\dagger}(k')] = \delta _{\mathbf{n} \mathbf{m} } \, \delta(k-k') , \qquad [ \hat{a} _{\mathbf{n}} (k) ,\hat{a} _{\mathbf{m}}(k')]=0. \label{ladder_algebra}
\end{align}
Here, $\hat{a}_{\mathbf{n}}(k)$ and $\hat{a}_{\mathbf{n}}^{\dagger}(k)$ denote the annihilation and creation operators, respectively. In this mode basis, the Hamiltonian operator becomes
\begin{align}
    \hat{H} = \sum_{\mathbf{n}} \int _{-\infty} ^{\infty} dk \; \hbar \, \omega _{\mathbf{n}} ^{(\lambda)} (k) \left[ \hat{a} _{\mathbf{n}} ^{\dagger}(k) \hat{a} _{\mathbf{n}} (k) + { \frac{L_z}{4\pi}} \right] , \label{Hamiltonian_separable}
\end{align}
so that the vacuum energy, $\langle 0 | \hat{H} | 0 \rangle$, is given by
\begin{align}
    E _{0} ^{(\lambda)} = { \frac{L_z}{4\pi}}\sum _{\mathbf{n}} \int _{-\infty} ^{\infty} dk \; \hbar \, \omega _{\mathbf{n}} ^{(\lambda)} (k) . \label{E0_formal_separable}
\end{align}

As can be seen from Eq. \eqref{E0_formal_separable}, the system is translationally invariant along the longitudinal direction $z$. 
As a consequence, the vacuum energy scales linearly with the corresponding length $L_z$, reflecting its extensive character. 
It is therefore natural to introduce the vacuum energy density per unit length, defined as
\begin{align}
    \mathcal{E} _{0} ^{(\lambda)} (L_{x},L_{y}) \equiv \lim _{L _{z} \to \infty } \frac{ E _{0} ^{(\lambda)}}{{ L _{z}}}. \label{E0_formal_z}
\end{align}
To evaluate this quantity, we temporarily introduce a finite quantization length $L _{z}$ along the longitudinal direction and impose periodic boundary conditions. In this case, the longitudinal momentum becomes discrete, $k = 2 \pi n / L _{z}$, and in the limit $L _{z} \to \infty$ the sum over $k$ is replaced by
\begin{align}
\sum _{k} \;\longrightarrow \; \frac{L _{z}}{2 \pi} \int _{-\infty} ^{\infty} dk .
\end{align}
Therefore, Eq. \eqref{E0_formal_z} takes the form
\begin{align}
    \mathcal{E} _{0} ^{(\lambda)} (L_{x},L_{y}) = \sum _{\mathbf{n}} \int _{-\infty}^{\infty} \frac{dk}{2 \pi} \; \frac{1}{2} \hbar \, \omega _{\mathbf{n}} ^{(\lambda )}(k). \label{E0_per_length_separable}
\end{align}
The expression of vacuum energy, \eqref{E0_per_length_separable}, is ultraviolet divergent and therefore requires regularization and renormalization. In the following subsection, we implement an appropriate regularization scheme and extract the finite Casimir energy for each configuration $(\lambda)$ by subtracting geometry-independent contributions. The resulting Casimir forces on the waveguide walls are then obtained from derivatives of the renormalized energy density with respect to $L_x$ and $L_y$.

\subsection{Casimir energy from twofold Abel-Plana} \label{twofold_AP_general_LV}

The vacuum energy density \eqref{E0_per_length_separable} is ultraviolet divergent, so it cannot be used directly as a well-defined starting point for algebraic manipulations of the discrete mode sums. In this work, however, we do not regularize the bare quantity and then renormalize it at the end. Instead, we implement renormalization from the outset by isolating the geometry-dependent interaction energy via a twofold ``sum minus integral'' prescription, which subtracts the bulk contribution and the local boundary self-energies.

Concretely, considering the transverse spectral function
\begin{align}
\omega _{n _{x}, n _{y}} ^{(\lambda)} (k) = c \, A ^{(\lambda)} \, \sqrt{ B ^{(\lambda)} \,  k ^{2} + C ^{(\lambda)} \, \left( \frac{n _{x} \pi}{L _{x}} \right) ^{2} + D ^{(\lambda)} \, \left( \frac{n _{y} \pi}{L _{y}} \right) ^{2} + \left( \frac{mc}{\hbar} \right) ^{2} } ,  \label{disp_general_case}
\end{align}
given by Eq. (\ref{disp_general_case}), we define the renormalized Casimir energy density per unit length along $z$ as
\begin{equation}
\mathcal{E}^{(\lambda)}_{C}(L_x,L_y)
=
\frac{\hbar}{2}
\int_{-\infty}^{\infty}\frac{dk}{2\pi}\,
\Bigg[
\sum_{n _{x}, n _{y} \ge 1} \omega _{n _{x}, n _{y}} ^{(\lambda)} (k) 
-
\int_{0}^{\infty}dn _{x}
\int_{0}^{\infty}dn _{y} \;
\omega ^{(\lambda)} (n _{x}, n _{y}; k)
\Bigg]_{\rm finite}.
\label{eq:EC_sum_minus_int_LV}
\end{equation}
Here $\omega _{n _{x}, n _{y}} ^{(\lambda)} (k)$ is the dispersion relation for the configuration $(\lambda)$, and $\omega^{(\lambda)}(n _{x}, n _{y};k)$ denotes its smooth continuation to real $(n _{x}, n _{y})$, obtained by replacing $n_{x} \mapsto x$ and $n _{y} \mapsto y$ in the mode frequencies. The subtraction in \eqref{eq:EC_sum_minus_int_LV} removes the geometry-independent bulk energy and the local ultraviolet pieces associated with the walls; the remaining finite part defines the physical Casimir interaction energy per unit length.

With this renormalized starting point, the discrete sums can be transformed in a controlled way by applying the Abel-Plana formula successively to $n _{x}$ and $n _{y}$. Importantly, the Abel-Plana remainders are finite because only branch-cut discontinuities contribute, while the polynomial (local) terms produced by the transformation precisely match the subtracted counterterms encoded in the integral piece of \eqref{eq:EC_sum_minus_int_LV}. In the following we therefore carry out a twofold Abel-Plana transformation of the bracket in \eqref{eq:EC_sum_minus_int_LV}, keeping track of the local contributions that cancel under the renormalization prescription and isolating the genuinely nonlocal (interaction) part that depends on $(L _{x} , L _{y} )$.

To evaluate the renormalized spectral sum appearing in \eqref{eq:EC_sum_minus_int_LV}, we employ the Abel-Plana summation formula \cite{Bordag:2009zz} successively in the two transverse directions:
\begin{align}
\sum_{n=1}^{\infty}F(n)
=
-\frac{1}{2}F(0)
+\int_{0}^{\infty}F(x)\,dx
+i\int_{0}^{\infty}dt\,
\frac{F(it)-F(-it)}{e^{2\pi t}-1},
\label{AP_identity}
\end{align}
first for the $n_x$ sum and then for the remaining $n_y$ sum. The successive application of the Abel-Plana formula decomposes the transverse spectral sum into three contributions: a local term, a continuum (integral) term, and a finite branch-cut remainder, defined explicitly in Eq.~\eqref{AP_identity}.

We now apply Eq.~\eqref{AP_identity} to the $n_x$ sum while keeping $n_y$ and $k$ fixed. For this purpose it is convenient to introduce the auxiliary function $\omega^{(\lambda)}(z,n_y;k)$, which denotes the analytic continuation of the dispersion relation under the complex extension of the transverse mode index $n_x\to z\in\mathbb{C}$, with $(z,n_y) \in \mathbb{C} \times \mathbb{N}$. Applying Eq.~\eqref{AP_identity} to the sum $\sum_{n_x=1}^{\infty}
\omega^{(\lambda)}_{n_x n_y}(k)$ yields
\begin{align}
\sum_{n_x=1} ^{\infty} \omega ^{(\lambda)} _{n_x n_y}(k) &= - \frac{1}{2} \, \omega ^{(\lambda)} (0,n _{y};k) + \int _{0} ^{\infty} dn_x\;\omega ^{(\lambda)}(n_x,n_y;k) + i\int_0^{\infty}dt\; \frac{ \omega^{(\lambda)}(it,n_y;k) - \omega^{(\lambda)}(-it,n_y;k) }{e ^{2\pi t}-1} . \label{AP_nx_LV_applied_new}
\end{align}
Substituting this result into the renormalized spectral sum
\eqref{eq:EC_sum_minus_int_LV}, the transverse contribution decomposes as
\begin{align}
\mathcal{S}^{(\lambda)} (k)
=
\mathcal{S}^{(\lambda)}_{\mathrm{loc}}(k)
+
\mathcal{S}^{(\lambda)}_{\mathrm{cont}}(k)
+
\mathcal{S}^{(\lambda)}_{\mathrm{AP}} (k),
\label{decomp_after_x_LV_new}
\end{align}
with
\begin{align}
\mathcal{S}^{(\lambda)}_{\mathrm{loc}} (k)
&=
-\frac12\sum_{n_y=1}^{\infty} 
\omega^{(\lambda)}(0,n_y;k),
\label{Ex0_LV_new}
\\[4pt]
\mathcal{S}^{(\lambda)}_{\mathrm{cont}} (k)
&=
\sum_{n_y=1}^{\infty}
\int_0^{\infty}dn_x\;
\omega^{(\lambda)}(n_x,n_y;k), \label{Ex_cont_LV_new}
\\[4pt]
\mathcal{S}^{(\lambda)}_{\mathrm{AP}} (k)
&=
i\sum_{n_y=1}^{\infty}
\int_0^{\infty}dt\;
\frac{
\omega^{(\lambda)}(it,n_y;k)
-
\omega^{(\lambda)}(-it,n_y;k)
}{e^{2\pi t}-1}.
\label{ExAP_LV_new}
\end{align}
These contributions arise, respectively, from the endpoint term of the analytic continuation of the $n_x$ spectrum, from the continuum part of the Abel-Plana representation, and from the Abel-Plana remainder, which captures the difference between the discrete transverse spectrum and its continuum limit.

In the following we present explicit closed-form expressions for each of these contributions. For clarity of exposition, only the final results are displayed in the main text, while the detailed derivations are collected in  Appendix~\ref{appendix_AP_derivations}. The first two contributions do not admit a significant further simplification, whereas the Abel-Plana remainder can be reduced to a compact form; for this reason, its derivation is discussed in greater detail in the Appendix. The result is
\begin{align}
\mathcal{S}^{(\lambda)}_{\mathrm{AP}} (k) &= \sum _{n _{y} = 1} ^{\infty}  g ^{(\lambda)} (n _{y},k) .  \label{SAP_main}
\end{align}
where
\begin{align}
g ^{(\lambda)} (n_{y}, k) &= - 2 c A ^{(\lambda)} \int _{t ^{(\lambda) } _{\ast}(n _{y}, k)} ^{\infty} dt \; \frac{ \sqrt{ C ^{(\lambda)} \big( \frac{\pi t}{L _{x} } \big) ^{2} - \Big[ B ^{(\lambda)} k ^{2} + D ^{(\lambda)} \big( \frac{\pi n _{y}}{L _{y}} \big) ^{2} + \big( \frac{mc}{\hbar} \big) ^{2} \Big] } }{ e ^{2 \pi t} - 1 } , \label{SAP_cut_main_text}
\end{align}
and
\begin{align}
t ^{(\lambda)} _{\ast} (n _{y} , k) = \frac{L _{x}}{\pi} \sqrt{ \frac{ B ^{(\lambda)} k ^{2} + D ^{(\lambda)} \big( \frac{\pi n _{y}}{L _{y}} \big) ^{2} + \big( \frac{mc}{\hbar} \big) ^{2} }{C ^{(\lambda)}} } , \label{t_threshold_correct}
\end{align}
which marks the onset of the branch-cut contribution. For $t < t ^{(\lambda)} _{\ast}(n _{y},k)$ no discontinuity is crossed and the Abel-Plana integrand vanishes, while for $t > t ^{(\lambda)} _{\ast}(n _{y},k)$ the square root becomes imaginary and produces a finite contribution.

At this stage, we proceed to apply the Abel-Plana formula once more, now to the remaining discrete sum over the transverse index $n_y$. This second transformation allows us to isolate the finite interaction part of the spectrum and complete the evaluation of the Casimir energy. 

For clarity of presentation, only the final expressions obtained after this second Abel-Plana transformation are reported in the main text. The detailed intermediate steps of the calculation are presented in Appendix~\ref{appendix_AP_derivations_B}, where the full analytic structure of the $n_y$ summation is worked out explicitly.

The local spectral contribution can be evaluated analytically by applying the Abel-Plana formula to the remaining discrete sum over the transverse index $n_y$. The calculation proceeds by analytically continuing the spectrum, identifying the associated branch cut, and separating the endpoint, continuum, and branch-cut contributions generated by the transformation. The resulting expression can be written in closed form as
\begin{align}
\mathcal{S}^{(\lambda)}_{\mathrm{loc}}(k)
=-
 \frac{cA^{(\lambda)}}{2}
\Bigg[-
 \frac{\Lambda^{(\lambda)}(k)}{2}
+
\frac{L_y}{\pi\sqrt{D^{(\lambda)}}}
\int_0^\infty d\chi\,
\sqrt{[\Lambda^{(\lambda)}(k)]^{2}+\chi^2}
-
2\int_{t^{(\lambda)}_{\ast\ast}(k)}^\infty dt\;
\frac{
\sqrt{
D^{(\lambda)}
\big(\frac{\pi t}{L_y}\big)^2
-
[\Lambda^{(\lambda)}(k)]^{2}
}
}{e^{2\pi t}-1}
\Bigg],
\end{align}
where
\begin{align}
t^{(\lambda)}_{\ast\ast}(k)
=
\frac{L_y}{\pi}
\frac{\Lambda^{(\lambda)}(k)}{\sqrt{D^{(\lambda)}}},
\qquad
\Lambda^{(\lambda)}(k)
=
\sqrt{B^{(\lambda)}k^2+\Big(\frac{mc}{\hbar}\Big)^2}.
\end{align}
Details of the derivation are presented in Appendix~\ref{appendix_AP_derivations_B}.

The continuum spectral contribution is obtained by applying the Abel-Plana formula to the remaining sum over the transverse index $n_y$, treating the integral over $n_x$ as part of the spectral density. The calculation proceeds by analytic continuation of the spectrum, identification of the branch cut, and evaluation of the associated discontinuity. After separating the endpoint, continuum, and branch-cut pieces, the result can be written as
\begin{align}
\mathcal{S}^{(\lambda)}_{\mathrm{cont}}(k)
=
- \frac{c A^{(\lambda)}}{2}  \left[ \frac{L_x}{\pi\sqrt{C^{(\lambda)}}}
\int_0^\infty d\chi\,
\sqrt{ [ \Lambda ^{(\lambda)} (k) ] ^{2} + \chi^2}
+ \frac{ L_x}{ \sqrt{C^{(\lambda)}}}
\int_{t^{(\lambda)}_{\ast\ast}(k)}^\infty dt
\frac{
D^{(\lambda)}\Big(\frac{\pi t}{L_y}\Big)^2
-
[ \Lambda ^{(\lambda)} (k) ] ^{2}
}{e^{2\pi t}-1} \right] \notag \\[5pt]
+ \int_0^\infty dn_y
\int_0^\infty dn_x \;
\omega^{(\lambda)}(n_x,n_y;k)  .
\end{align}
Details of the derivation are presented in Appendix~\ref{appendix_AP_derivations_B}.

The remaining contribution arises from the second Abel-Plana transformation applied to the branch-cut term generated by the first transverse summation. Its evaluation is controlled by the analytic continuation of the dispersion relation under $n_y\to z$.

When the continuation $z=\pm it$ is performed, the square-root structure of the spectrum becomes sign-indefinite. Above a threshold value of $t$, the argument of the square root becomes negative and the spectral function develops a branch cut in the complex plane. The Abel-Plana remainder is then entirely determined by the discontinuity across this cut. The detailed derivation is presented in Appendix~\ref{appendix_AP_derivations}; here we quote the final result.

The second Abel-Plana transformation gives
\begin{align}
\sum_{n_y=1}^{\infty} g^{(\lambda)}(n_y,k)
&=
c A ^{(\lambda)}
\int_{t^{(\lambda)}_{\ast}(0,k)}^\infty dt\;
\frac{
\sqrt{
C^{(\lambda)}\Big(\frac{\pi t}{L_x}\Big)^2
-
[ \Lambda ^{(\lambda)} (k) ] ^{2}
}
}{e^{2\pi t}-1}
-  
\frac{ c A ^{(\lambda)} L_y}{2\sqrt{D^{(\lambda)}}}
\int_{t^{(\lambda)}_{\ast}(0,k)}^\infty dt\;
\frac{
C^{(\lambda)}\Big(\frac{\pi t}{L_x}\Big)^2
-
[ \Lambda ^{(\lambda)} (k) ] ^{2}
}{e^{2\pi t}-1} \notag \\ & \hspace{3cm}
+i\int_{t^{(\lambda)}_{\ast\ast}(k)}^\infty dt\;
\frac{g^{(\lambda)}(it,k)-g^{(\lambda)}(-it,k)}{e^{2\pi t}-1}.
\end{align}
The discontinuity across the branch cut is given by
\begin{align}
g^{(\lambda)}(it,k)-g^{(\lambda)}(-it,k)
= -   
\frac{4i c A ^{(\lambda)} L_x}{\pi\sqrt{C^{(\lambda)}}}
\int_{-B^{(\lambda)}(t,k)}^{B^{(\lambda)}(t,k)} ds\;
\sqrt{ [B^{(\lambda)}(t,k)] ^2-s^2}
\frac{1}{e^{2iL_x s/\sqrt{C^{(\lambda)}}}-1},
\qquad t>t^{(\lambda)}_{\ast\ast}(k) ,
\end{align}
where
\begin{align}
B^{(\lambda)}(t,k)
=
\sqrt{
D^{(\lambda)}\Big(\frac{\pi t}{L_y}\Big)^2
-
[ \Lambda ^{(\lambda)} (k) ] ^{2} } . 
\end{align}
Thus, the nonlocal part of the transverse spectral sum is completely governed by the integral of the spectral kernel along the branch cut in the auxiliary complex plane.

Collecting the results of the twofold Abel-Plana procedure, and symmetrizing the second Abel-Plana transformation of the branch-cut contribution to ensure independence of the summation order, the renormalized transverse spectral bracket at fixed $k$ can be written compactly as
\begin{align}
\sum_{n_x,n_y\ge 1}\omega^{(\lambda)}_{n_x n_y}(k)
-\int_0^\infty dn_x\int_0^\infty dn_y\;\omega^{(\lambda)}(n_x,n_y;k) &= \frac{cA^{(\lambda)} \Lambda^{(\lambda)}(k)}{4} - cA^{(\lambda)} \,
\tilde{L} _{x} \,  \mathcal{I} (k)
+  c A^{(\lambda)} \,   \mathcal{J} (k, \tilde{L} _{x})  
\notag \\ & 
+ c A ^{(\lambda)}  [ \mathcal{M} (k,\tilde{L} _{y},\tilde{L} _{x})  -  \tilde{L} _{x} \,   \mathcal{K} (k, \tilde{L} _{y})   ]       + (\tilde{L} _{y} \leftrightarrow \tilde{L} _{x})  ,  \label{regularized_sum}
\end{align}
where we have defined the integral functions
\begin{align}
    \mathcal{I} (k) &= \frac{1}{2 \pi} \int_0^\infty d\chi\,
\sqrt{ [ \Lambda ^{(\lambda)} (k) ] ^{2} + \chi^2} , \label{I_function} \\  \mathcal{J} (k,L) &= \int_{  L \Lambda^{(\lambda)}(k) / \pi }^\infty dt\;
\frac{B ^{(\lambda)} (t,k,L)
}{e^{2\pi t}-1}  \label{J_function} \\  \mathcal{K} (k,L) &= \frac{1}{2} \int_{  L \Lambda^{(\lambda)}(k) / \pi }^\infty dt\;
\frac{ 
[B ^{(\lambda)} (t,k,L)]^{2}
}{e^{2\pi t}-1}  \label{K_function} , \\ \mathcal{M} (k,L,L') &= \frac{4L}{\pi} \; \int_{  L' \Lambda^{(\lambda)}(k) / \pi }^\infty dt\;
\frac{1}{e^{2\pi t}-1}    
\int_{- B ^{(\lambda)} (t,k,L') } ^{ + B ^{(\lambda)} (t,k,L')  } ds\;
\frac{\sqrt{ [ B ^{(\lambda)} (t,k,L)] ^{2} - s ^{2} }}{e^{2i L s}-1}  , \label{M_function}
\end{align}
and the rescaled lengths $\tilde{L} _{x} = L _{x} / \sqrt{C^{(\lambda)}} $ and $\tilde{L} _{y} = L _{y} / \sqrt{D^{(\lambda)}} $. Here, $B ^{(\lambda)} (t,k,L) = \sqrt{ \big( \pi t / L \big)^2 - [ \Lambda ^{(\lambda)} (k) ] ^{2} } $.

The first two contributions on the right-hand side of Eq. \eqref{regularized_sum}, $cA^{(\lambda)}\Lambda^{(\lambda)}(k) / 4$ and $- cA^{(\lambda)}\tilde L_x\,\mathcal I(k)$ (together with the symmetric $\tilde L_y$ piece), are purely local ultraviolet terms. In particular, $\mathcal I(k)$ is linearly divergent and multiplies the rescaled boundary length, so it represents a geometry-independent bulk/self-energy contribution (and a local surface energy proportional to the wall length), rather than an interaction between distinct boundaries. Such terms do not encode measurable Casimir forces and are removed by the renormalization prescription implicit in the ``sum minus integral'' definition, i.e. by fixing the vacuum energy density and boundary self-energies to their reference values. Moreover, the combination $\mathcal{M}(k,\tilde L_y,\tilde L_x) - \tilde L_x\,\mathcal{K}(k,\tilde L_y)$ is finite. The term $\tilde L_x\,\mathcal{K}(k,\tilde L_y)$ represents a local contribution proportional to the boundary length, which is precisely matched by the local part contained in $\mathcal{M}(k,\tilde L_y,\tilde L_x)$. These local pieces cancel identically, leaving only the genuinely nonlocal, finite part of $\mathcal{M}(k,\tilde L_y,\tilde L_x)$, which encodes the interaction between distinct boundaries. This cancellation is demonstrated explicitly in the following. To separate local and nonlocal contributions we use the exact identity
\begin{equation}
\frac{1}{e^{2ix}-1} = -\frac{1}{2} - \frac{i}{2} \cot x ,
\end{equation}
which splits the cut kernel into a purely local part and a finite nonlocal part.  Substituting this decomposition into Eq. \eqref{M_function} yields a decomposition $\mathcal{M} (k,L,L') = \mathcal{M} _{\rm loc} (k,L,L') + \mathcal{M} _{\rm fin} (k,L,L')$, where
\begin{align}
    \mathcal{M} _{\rm loc} (k,L,L') &= - \frac{2L}{\pi} \; \int_{  L' \Lambda^{(\lambda)}(k) / \pi }^\infty dt\;
\frac{1}{e^{2\pi t}-1}    
\int_{- B ^{(\lambda)} (t,k,L') } ^{ + B ^{(\lambda)} (t,k,L')  } ds\; \sqrt{ [ B ^{(\lambda)} (t,k,L)] ^{2} - s ^{2} } = L \, \mathcal{K} (k,L ') ,  
\end{align}
which exactly cancels the explicit local contribution proportional to $- L \, \mathcal{K} (k,L ')$. Only the finite nonlocal part therefore survives. The finite contribution originates from the cotangent term and reads
\begin{align}
    \mathcal{M} _{\rm fin} (k,L,L') &= \frac{2L}{i \pi} \; \int_{  L' \Lambda^{(\lambda)}(k) / \pi }^\infty dt\;
\frac{1}{e^{2\pi t}-1}    
\int_{- B ^{(\lambda)} (t,k,L') } ^{ + B ^{(\lambda)} (t,k,L')  } ds\; \sqrt{ [ B ^{(\lambda)} (t,k,L)] ^{2} - s ^{2} } \, \cot (L s)  . 
\end{align}
The integral involving $\cot (L s)$ is understood in the sense of the Cauchy principal value. This prescription is required because the cotangent has simple poles  on the real axis at $s=n\pi/L$, which may lie within the integration interval.  The principal value arises naturally from the symmetric evaluation of the branch-cut discontinuity in the complex plane and ensures that only the nonlocal part of the spectral kernel contributes to the physical result.

All of the above implies that the renormalized Casimir energy density per unit length along $z$ is given by
\begin{equation}
\mathcal{E}^{(\lambda)}_{C}(L_x,L_y)
=
\frac{\hbar c }{2}  A^{(\lambda)} 
\int_{-\infty}^{\infty}\frac{dk}{2\pi}\,
\Bigg[   \mathcal{J} (k, \tilde{L} _{x})   +  \mathcal{M} _{\rm fin} (k,\tilde{L} _{x} , \tilde{L} _{y} ) +   (\tilde{L} _{y} \leftrightarrow \tilde{L} _{x}) \Bigg] . \label{eq:EC_sum_minus_int_LV_fin}
\end{equation}
The finite contributions obtained after the second application of the Abel-Plana formula can be expressed in a more compact and transparent form by expanding the Bose-Einstein factor as a convergent exponential series,
\begin{align}
\frac{1}{e ^{2 \pi v}-1} = \sum _{n=1} ^{\infty} e ^{-2 \pi n v}.
\end{align}
This representation allows the remaining integrals over the auxiliary variables to be evaluated analytically in terms of modified Bessel functions and elementary exponential integrals, leading to closed-form expressions for each of the sectors generated by the twofold Abel-Plana transformation. Using the above expansion the expression \eqref{J_function} for the function $\mathcal{J}$ becomes
\begin{align}
\mathcal{J}(k,\tilde{L} _{x})
&=
\frac{\tilde{L} _{x}}{\pi}\sum_{n=1}^{\infty}
\int_{\Lambda^{(\lambda)}(k)}^{\infty}du\;
\sqrt{u^{2}-\big[\Lambda^{(\lambda)}(k)\big]^2}\;
e^{-2n \tilde{L} _{x} u}.
\label{eq:J_after_u_change}
\end{align}
The remaining integral is standard and can be expressed in terms of a modified Bessel \cite{Abra} function,
\begin{equation}
\int_{\mu }^{\infty}du\;e^{-\beta u}\sqrt{u^{2}-\mu ^{2}}
=
\frac{\mu}{\beta}\,K_{1}(\mu \beta),
\qquad \Re(\beta)>0,\ \Re(\mu )>0.
\label{eq:Bessel_identity_J}
\end{equation}
With $\mu = \Lambda^{(\lambda)}(k)$ and $\beta=2n \tilde{L} _{x} $, Eq.~\eqref{eq:J_after_u_change}
gives the exponentially convergent series
\begin{equation}
\mathcal{J}(k,\tilde{L} _{x} ) = \frac{\Lambda ^{(\lambda)}(k)}{2 \pi} \sum _{n = 1} ^{\infty} \frac{1}{n} \, K _{1} \Big( 2 n \tilde{L} _{x} \, \Lambda ^{(\lambda)} (k) \Big) . \label{eq:J_Bessel_series}
\end{equation}
To evaluate the finite contribution $\mathcal{M}_{\rm fin}(k,L,L')$ we proceed in the same manner as for $\mathcal{J}$, using the exponential representation of the Bose-Einstein factor introduced above. The $\cot(Ls)$ kernel is treated through its Mittag-Leffler expansion,
\begin{equation}
\cot(Ls)=\frac{1}{Ls}
+2Ls\sum_{p=1}^{\infty}\frac{1}{(Ls)^2-\pi^2 p^2}.
\label{eq:cot_ML_clean}
\end{equation}
Since the singular term $1/(Ls)$ corresponds to the local contribution already isolated in $\mathcal{M}_{\rm loc}$, only the regular part of the Mittag-Leffler expansion contributes to $\mathcal{M}_{\rm fin}$. Because $\cot(Ls)$ has simple poles at $s=p\pi/L$, the $s$-integration is understood in the Cauchy principal-value sense, which follows from the symmetric evaluation of the branch-cut discontinuity in the complex plane.

Substituting the regular part of \eqref{eq:cot_ML_clean} into $\mathcal{M}_{\rm fin}$ and using the exponential series for the Bose-Einstein factor introduced above, we obtain
\begin{align}
\mathcal{M}_{\rm fin}(k,L,L')
&=
\frac{2L}{i\pi}
\sum_{q=1}^{\infty}
\int_{L'\Lambda^{(\lambda)}(k)/\pi}^{\infty} dt\;e^{-2\pi q t}\;
\mathrm{PV} \int_{-B^{(\lambda)}(t,k,L')}^{+B^{(\lambda)}(t,k,L')} ds\;
\sqrt{\big[B^{(\lambda)}(t,k,L)\big]^2-s^2}\,
\Bigg(2Ls\sum_{p=1}^{\infty}\frac{1}{(Ls)^2-\pi^2 p^2}\Bigg),
\label{eq:Mfin_after_expansions}
\end{align}
where $B^{(\lambda)}(t,k,L)=\sqrt{\big(\pi t/L\big)^2-[\Lambda^{(\lambda)}(k)]^2}$.
Interchanging the $p$-sum with the principal-value $s$-integral, the latter reduces to the elementary cut integral
\begin{equation}
\mathrm{PV} \int_{-B}^{B} ds\;
\frac{s\,\sqrt{B^2-s^2}}{(Ls)^2-\pi^2 p^2}
=
-\frac{\pi}{2L^2}\,
\sqrt{B^2+\Big(\frac{\pi p}{L}\Big)^2}
+\frac{\pi^2 p}{2L^3},
\qquad (p\ge1),
\label{eq:PV_cut_integral_key}
\end{equation}
so that \eqref{eq:Mfin_after_expansions} becomes a single $t$-integral with a square-root kernel. Introducing the variable
\begin{equation}
u=\frac{\pi t}{L'},\qquad
t=\frac{L'}{\pi}u,\qquad
B^{(\lambda)}(t,k,L')=\sqrt{u^2-[\Lambda^{(\lambda)}(k)]^2},
\label{eq:u_change_Mfin}
\end{equation}
one obtains terms of the generic form
\begin{equation}
\int_{\Lambda^{(\lambda)}(k)}^{\infty} du\;
e^{-2qL' u}\,
\sqrt{u^2-[\Lambda^{(\lambda)}(k)]^2}
=
\frac{\Lambda^{(\lambda)}(k)}{2qL'}\,
K_{1} \Big(2qL'\Lambda^{(\lambda)}(k)\Big),
\label{eq:K1_identity_Mfin}
\end{equation}
which is the same standard Bessel integral used previously for $\mathcal{J}$. Collecting all pieces, the finite mixed contribution can be written as the exponentially convergent double series
\begin{equation}
\mathcal{M}_{\rm fin}(k,L,L')
=
-\frac{\Lambda^{(\lambda)}(k)}{\pi}
\sum_{p=1}^{\infty}\sum_{q=1}^{\infty}
\frac{K_{1}\!\Big(2\Lambda^{(\lambda)}(k)\,
\sqrt{(pL)^2+(qL')^2}\Big)}
{\sqrt{(pL)^2+(qL')^2}}, \label{eq:Mfin_doubleK1_clean}
\end{equation}
which is manifestly finite and symmetric under $L\leftrightarrow L'$.

Substituting the above results into Eq. \eqref{eq:EC_sum_minus_int_LV_fin} we find
\begin{align}
\mathcal{E}^{(\lambda)}_{C}(L_x,L_y)
&=
\frac{\hbar c }{4 \pi}  A^{(\lambda)} 
\int_{-\infty}^{\infty}\frac{dk}{2\pi}\, \Lambda ^{(\lambda)}(k)  
\Bigg[  \sum _{p = 1} ^{\infty} \frac{1}{p} \, K _{1} \Big( 2 p \tilde{L} _{x} \, \Lambda ^{(\lambda)} (k) \Big) +    \sum _{p = 1} ^{\infty} \frac{1}{p} \, K _{1} \Big( 2 p \tilde{L} _{y} \, \Lambda ^{(\lambda)} (k) \Big) \notag \\ & \hspace{6cm} - 2   
\sum_{p=1}^{\infty}\sum_{q=1}^{\infty}
\frac{K_{1} \Big(2\Lambda^{(\lambda)}(k)\,
\sqrt{(p \tilde{L} _{x}  )^2+(q \tilde{L} _{y} )^2}\Big)}
{\sqrt{(p \tilde{L} _{x}  )^2+(q \tilde{L} _{y} )^2}} \Bigg] . \label{eq:EC_sum_minus_int_LV_fin2}
\end{align}
The remaining integral over the longitudinal momentum $k$ can be carried out analytically. Since the integrand depends on $k$ only through
\begin{equation}
\Lambda^{(\lambda)}(k)
=
\sqrt{B^{(\lambda)}k^2+\mu^2},
\qquad
\mu\equiv\frac{mc}{\hbar},
\end{equation}
and is even in $k$, we write
\begin{equation}
\int_{-\infty}^{\infty}\frac{dk}{2\pi}\,
\Lambda^{(\lambda)}(k)\,F \big(\Lambda^{(\lambda)}(k)\big)
=
\frac{1}{\pi}\int_{0}^{\infty}dk\,
\Lambda^{(\lambda)}(k)\,F \big(\Lambda^{(\lambda)}(k)\big).
\end{equation}
Introducing the standard hyperbolic parametrization
\begin{equation}
k=\frac{\mu}{\sqrt{B^{(\lambda)}}}\sinh u,
\qquad
\Lambda^{(\lambda)}(k)=\mu\cosh u,
\qquad
dk=\frac{\mu}{\sqrt{B^{(\lambda)}}}\cosh u\,du,
\end{equation}
the integral becomes
\begin{equation}
\int_{-\infty}^{\infty}\frac{dk}{2\pi}\,
\Lambda^{(\lambda)}(k)\,F \big(\Lambda^{(\lambda)}(k)\big)
=
\frac{\mu^2}{\pi\sqrt{B^{(\lambda)}}}
\int_{0}^{\infty}du\,
\cosh^2 u\,
F(\mu\cosh u).
\label{eq:k_integral_general}
\end{equation}
For the functions appearing in Eq.~\eqref{eq:EC_sum_minus_int_LV_fin2}, one needs the identity
\begin{equation}
\int_{0}^{\infty}du\,
\cosh^2 u\,
K_1(a\mu\cosh u)
=
\frac{1}{a\mu}\,K_2(a\mu),
\qquad a>0,
\label{eq:K_identity_used}
\end{equation}
which follows from standard recurrence relations of the modified Bessel functions. Using this result, we obtain
\begin{equation}
\int_{-\infty}^{\infty}\frac{dk}{2\pi}\,
\Lambda^{(\lambda)}(k)\,
K_1 \big(a\,\Lambda^{(\lambda)}(k)\big)
=
\frac{\mu^2}{\pi\sqrt{B^{(\lambda)}}}\,
\frac{1}{a}\,
K_2 (a\mu).
\label{eq:k_integral_master}
\end{equation} 
Applying Eq.~\eqref{eq:k_integral_master} to the three sectors of the Casimir energy expression, with
\begin{align}
    a=2p\tilde L_x,
\qquad
a=2p\tilde L_y,
\qquad
a=2R_{pq},
\qquad
R_{pq}=\sqrt{(p\tilde L_x)^2+(q\tilde L_y)^2},
\end{align}
we finally obtain the closed form
\begin{align}
\mathcal{E}^{(\lambda)}_{C}(L_x,L_y)
&=
\frac{\hbar c\,A^{(\lambda)}\,\mu^2}
{8\pi^2\sqrt{B^{(\lambda)}}}
\Bigg[
\sum_{p=1}^{\infty}
\frac{K_2 \big(2p\tilde L_x\,\mu\big)}
{p^2\,\tilde L_x}
+
\sum_{p=1}^{\infty}
\frac{K_2 \big(2p\tilde L_y\,\mu\big)}
{p^2\,\tilde L_y}
-
\sum_{p,q\ge1}
\frac{
K_2 \big(2\mu R_{pq}\big)
}{
R_{pq}
}
\Bigg],
\label{eq:Casimir_final_compact}
\end{align}
with $\mu=mc/\hbar$.

\begin{figure}[h!]
  \centering
  \begin{subfigure}{0.45\textwidth}
    \centering
    \includegraphics[width=\linewidth]{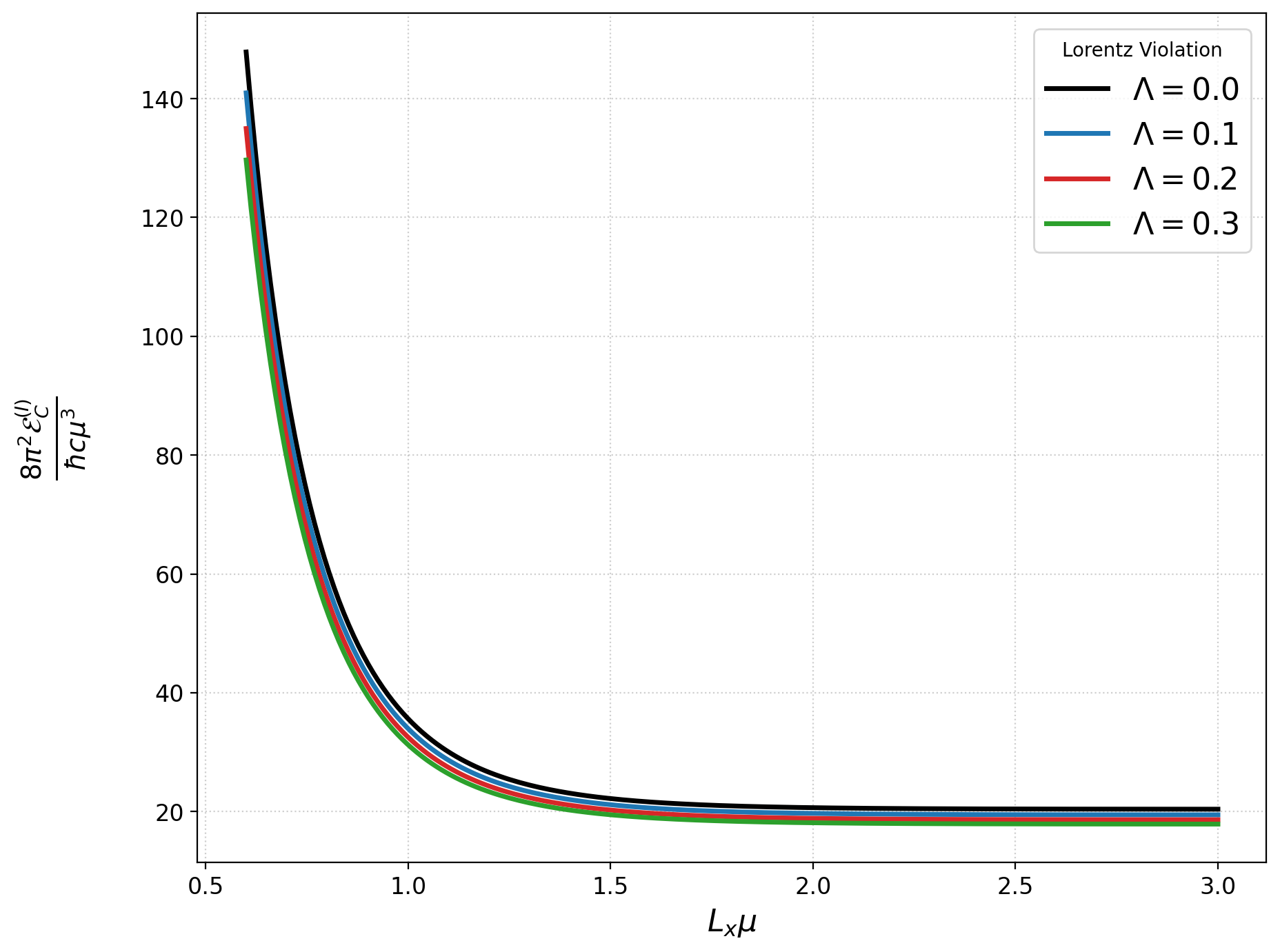}
  \end{subfigure}
  \begin{subfigure}{0.45\textwidth}
    \centering
    \includegraphics[width=\linewidth]{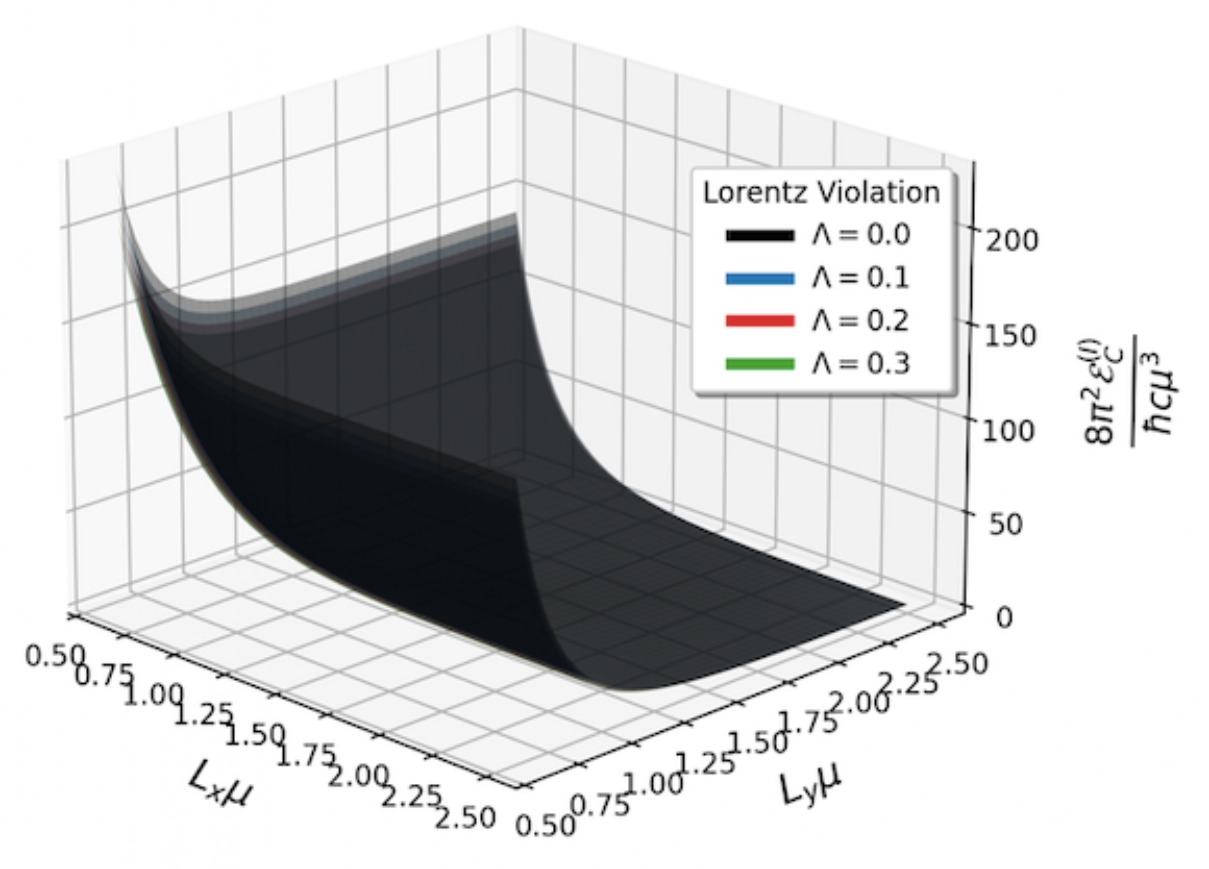}
  \end{subfigure}
  \caption{Dimensionless Casimir energy density $8 \pi^2 \mathcal{E}_C^{(I)}/\hbar c \mu^3$ for Case I as a function of the plate separation distances $L_x\mu$ and $L_y\mu$ for different Lorentz-violating parameters $\Lambda = \{0.0, 0.1, 0.2, 0.3\}$. The left panel shows the 1D energy profile for a fixed $L_y\mu = 1.0$, highlighting the attenuation of the vacuum energy as $\Lambda$ increases. The right panel displays the 3D overlapping energy surfaces, providing a global perspective of the energy decay.}
  \label{casimir_I_ab}
\end{figure}

\begin{figure}[h!]
    \centering
    \includegraphics[width=0.85\linewidth]{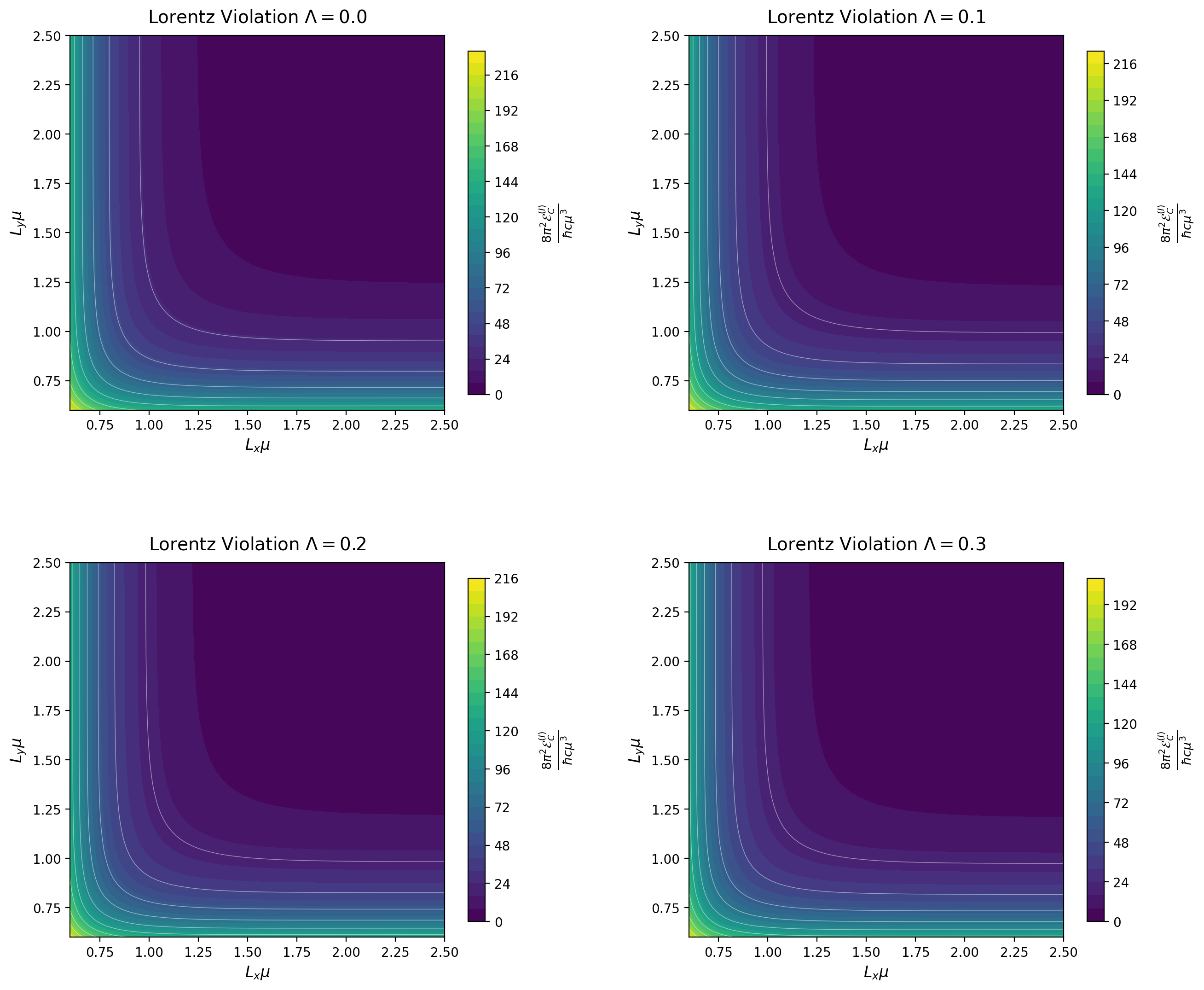}
    \caption{Contour maps of the Casimir energy density for Case I, considering four distinct values of the LV parameter: $\Lambda = 0.0$ (standard Case), $0.1, 0.2$ and $0.3$. The color gradient and the superimposed iso-energy lines represent the magnitude of $8 \pi^2 \mathcal{E}_C^{(I)}/\hbar c \mu^3$. The perfect symmetry of the concentric contours across all panels demonstrates that Case I preserves spatial isotropy in the $L_x$-$L_y$ plane, while the LV parameter acts as a global scaling factor for the energy intensity.}
    \label{casimir_I_c}
\end{figure}

\begin{figure}[h!]
  \centering
  \begin{subfigure}{0.45\textwidth}
    \centering
    \includegraphics[width=\linewidth]{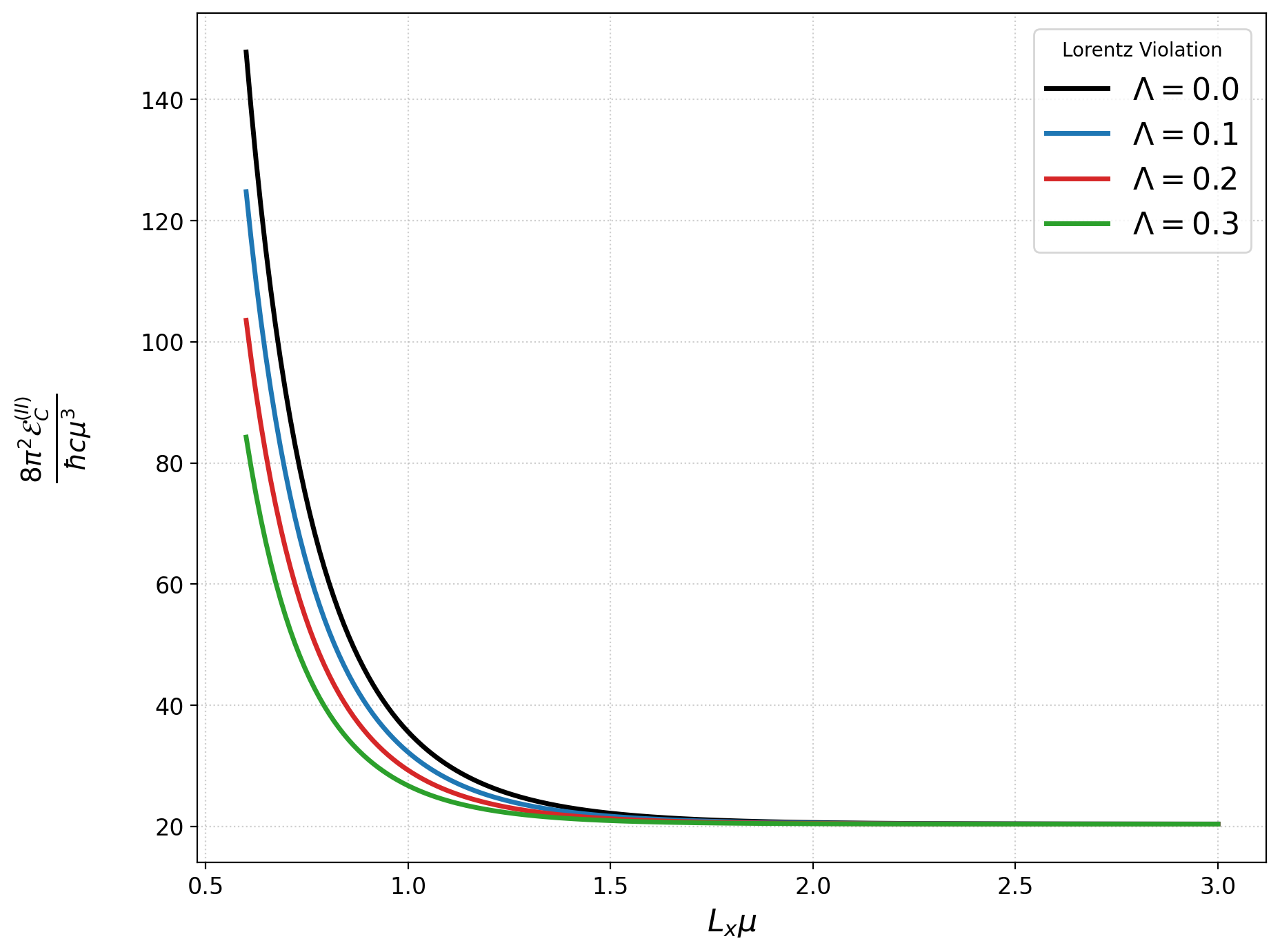}
  \end{subfigure}
  \begin{subfigure}{0.45\textwidth}
    \centering
    \includegraphics[width=\linewidth]{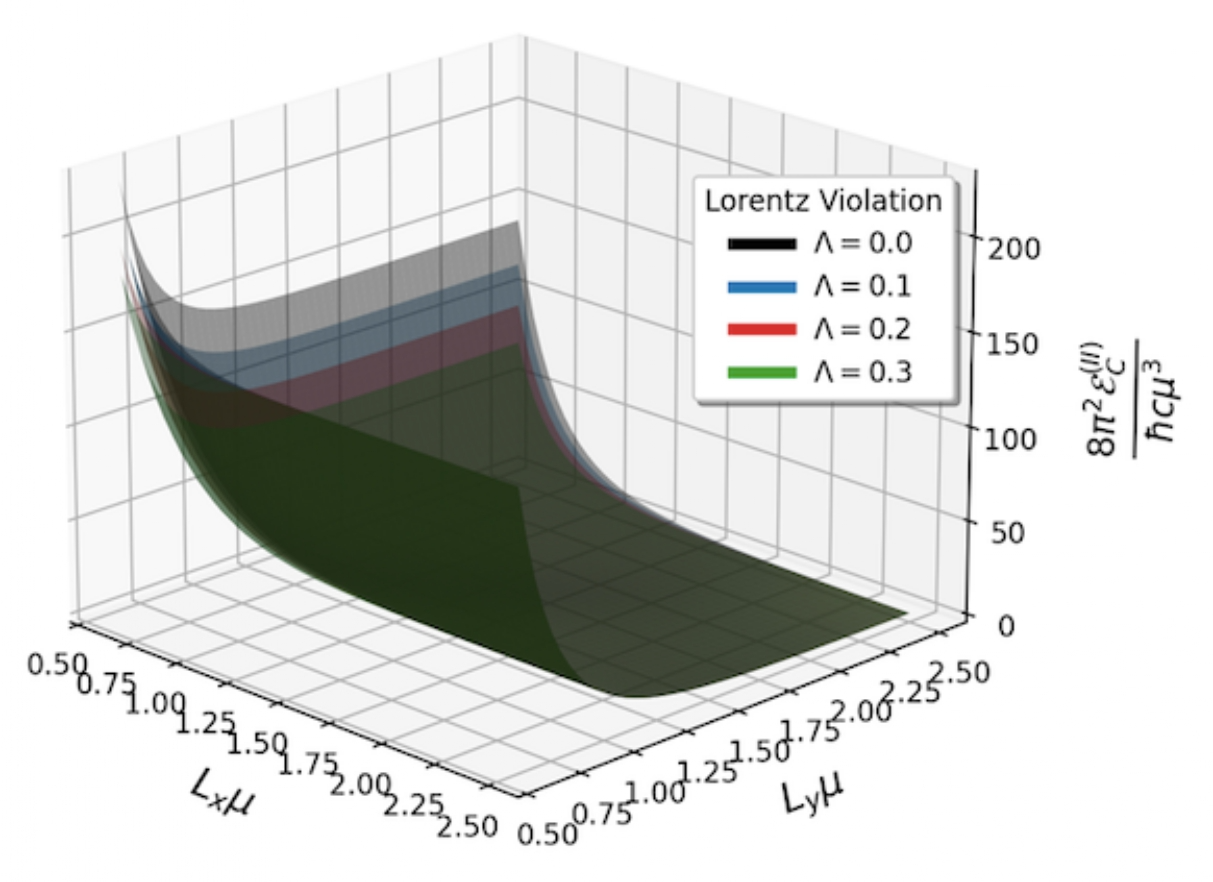}
  \end{subfigure}
  \caption{Dimensionless Casimir energy density $8 \pi^2 \mathcal{E}_C^{(II)}/\hbar c \mu^3$ for Case II as a function of the plate separation distances $L_x\mu$ and $L_y\mu$ for different Lorentz-violating parameters $\Lambda = \{0.0, 0.1, 0.2, 0.3\}$. The left panel shows the 1D energy profile for a fixed $L_y\mu = 1.0$, highlighting the anisotropic attenuation of the vacuum energy magnitude as $\Lambda$ increases. The right panel displays the 3D overlapping energy surfaces, providing a global perspective of the energy decay and the geometric deformation induced by the spatial Lorentz-violating component $u_1$.}
  \label{casimir_II_ab}
\end{figure}

\begin{figure}[h!]
    \centering
    \includegraphics[width=0.85\linewidth]{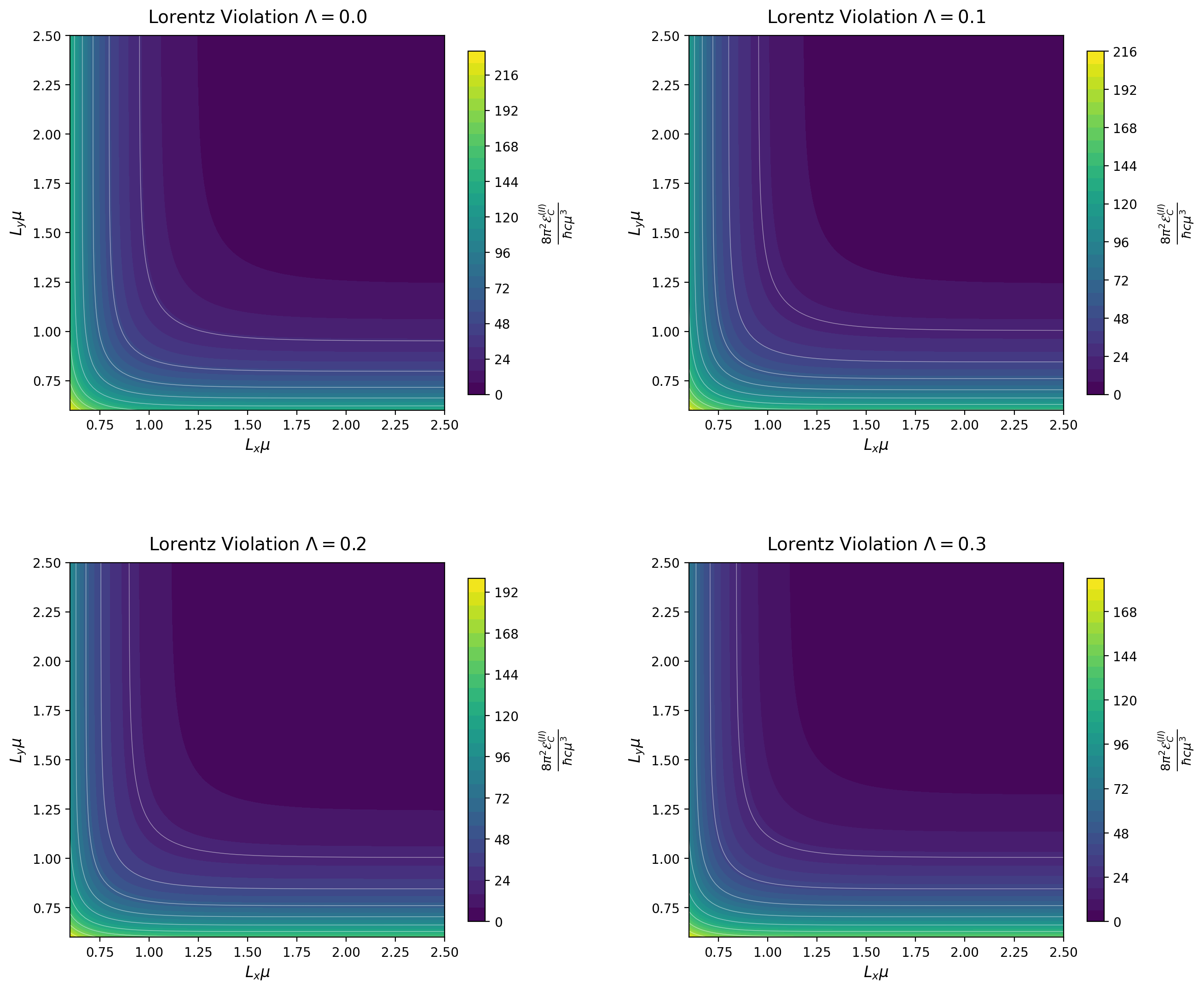}
    \caption{Contour maps of the Casimir energy density for Case II, considering four distinct values of the LV parameter: $\Lambda = 0.0$ (standard Case), $0.1, 0.2,$ and $0.3$. The color gradient and the superimposed iso-energy lines represent the magnitude of $8 \pi^2 \mathcal{E}_C^{(II)}/\hbar c \mu^3$. The progressive elongation of the contours along the $L_x \mu$ axis demonstrates that Case II breaks spatial isotropy in the $L_x$-$L_y$ plane, as the LV parameter introduces a directional deformation in the effective confinement scale.}
    \label{casimir_II_c}
\end{figure}

\begin{figure}[h!]
  \centering
  \begin{subfigure}{0.45\textwidth}
    \centering
    \includegraphics[width=\linewidth]{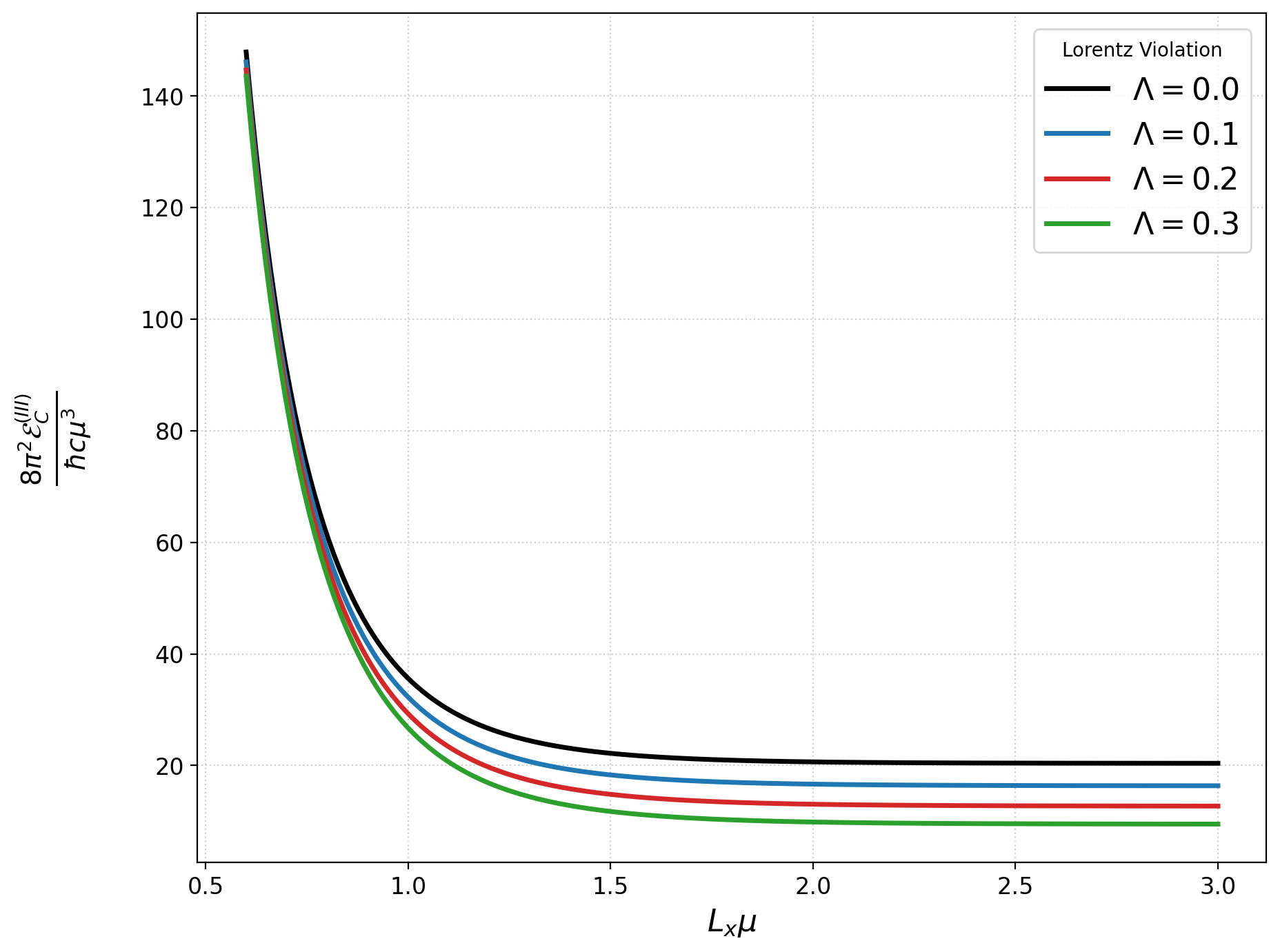}
  \end{subfigure}
  \begin{subfigure}{0.45\textwidth}
    \centering
    \includegraphics[width=\linewidth]{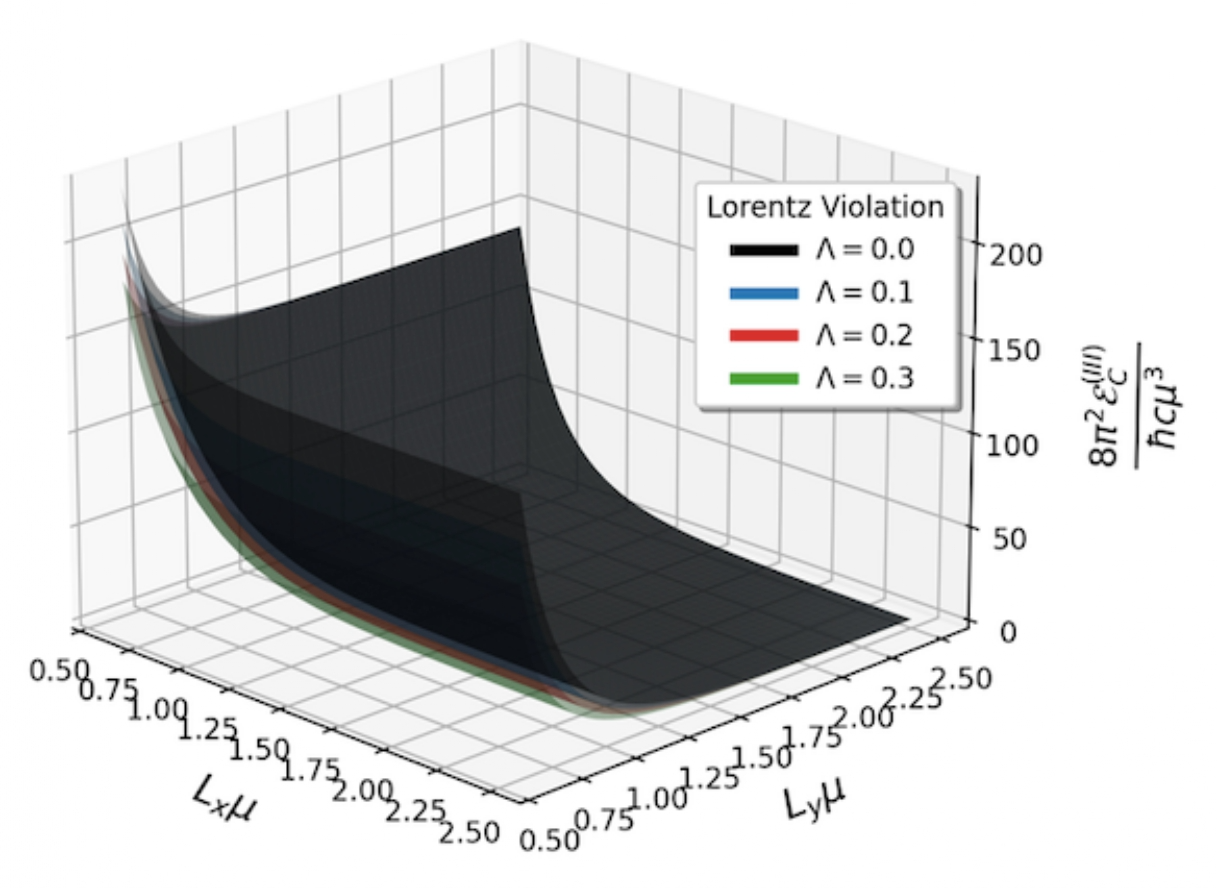}
  \end{subfigure}
  \caption{Dimensionless Casimir energy density $8 \pi^2 \mathcal{E}_C^{(III)}/\hbar c \mu^3$ for Case III as a function of the plate separation distances $L_x\mu$ and $L_y\mu$ for different Lorentz-violating parameters $\Lambda = \{0.0, 0.1, 0.2, 0.3\}$. The left panel shows the 1D energy profile for a fixed $L_y\mu = 1.0$, highlighting the anisotropic shift in vacuum energy as $\Lambda$ increases. The right panel displays the 3D overlapping energy surfaces, providing a global perspective of the energy decay and the transverse asymmetry induced by the spatial LV component $u_2$ along the $L_y$ direction.}
  \label{casimir_III_ab}
\end{figure}

\begin{figure}[h!]
    \centering
    \includegraphics[width=0.85\linewidth]{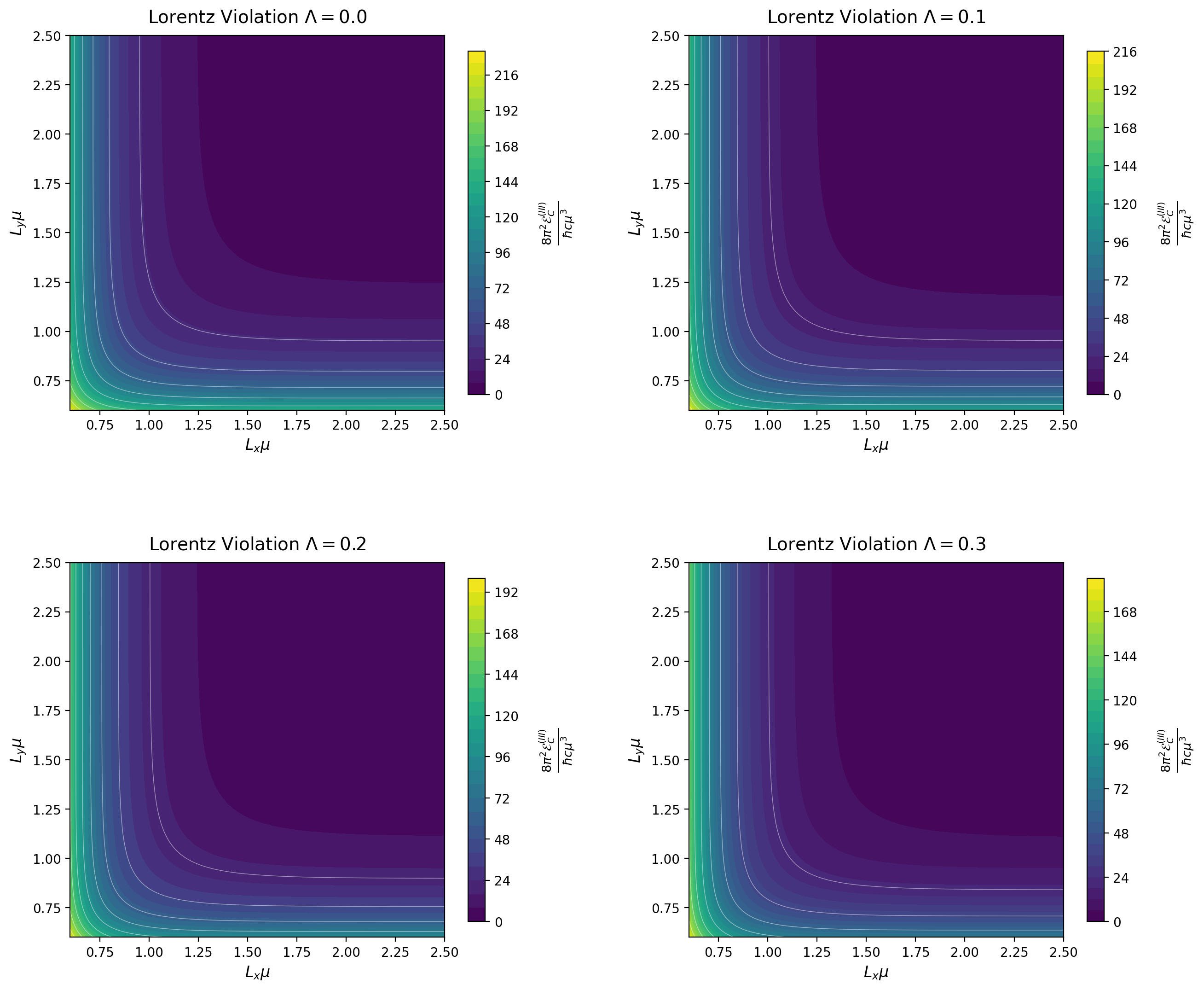}
    \caption{Contour maps of the Casimir energy density for Case III, considering four distinct values of the LV parameter: $\Lambda = 0.0$ (standard Case), $0.1, 0.2,$ and $0.3$. The color gradient and the superimposed iso-energy lines represent the magnitude of $8 \pi^2 \mathcal{E}_C^{(III)}/\hbar c \mu^3$. The progressive vertical elongation of the contours along the $L_y \mu$ axis demonstrates that Case III breaks spatial isotropy in the $L_x$-$L_y$ plane, as the LV parameter $u_2$ induces a directional deformation in the effective transverse confinement scale.}
    \label{casimir_III_c}
\end{figure}

\begin{figure}[h!]
  \centering
  \begin{subfigure}{0.45\textwidth}
    \centering
    \includegraphics[width=\linewidth]{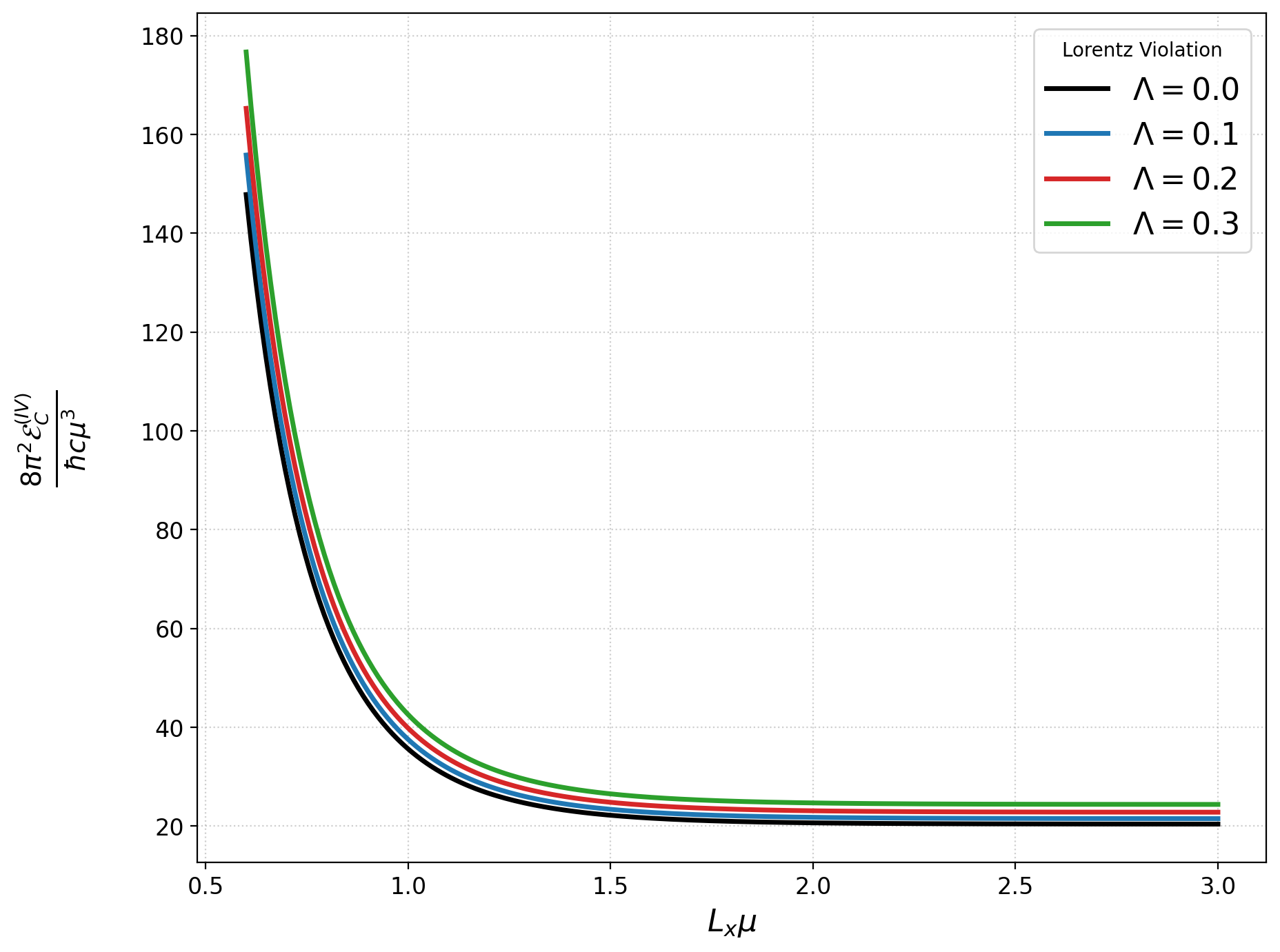}
  \end{subfigure}
  \begin{subfigure}{0.45\textwidth}
    \centering
    \includegraphics[width=\linewidth]{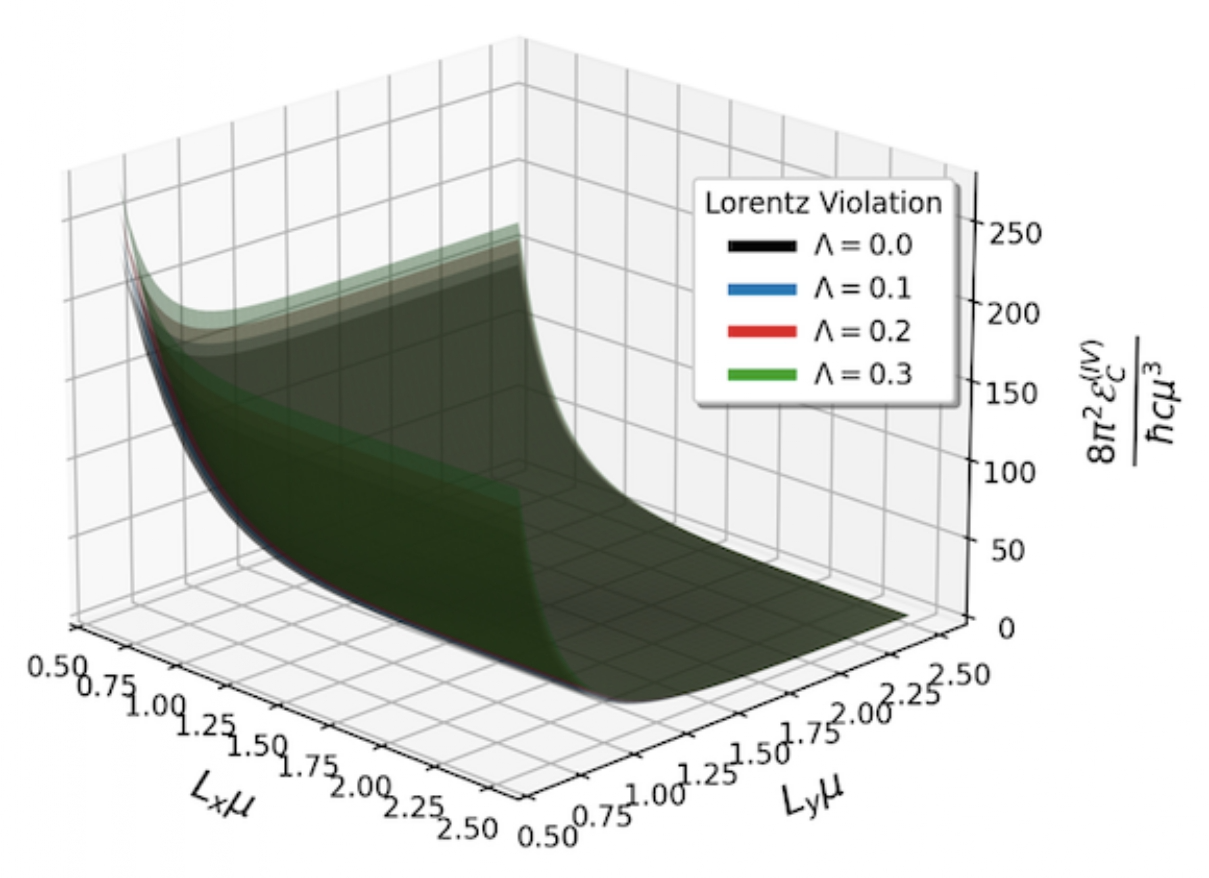}
  \end{subfigure}
  \caption{Dimensionless Casimir energy density $8 \pi^2 \mathcal{E}_C^{(IV)}/\hbar c \mu^3$ for Case IV as a function of the plate separation distances $L_x\mu$ and $L_y\mu$ for different Lorentz-violating parameters $\Lambda = \{0.0, 0.1, 0.2, 0.3\}$. The left panel shows the 1D energy profile for a fixed $L_y\mu = 1.0$, highlighting the enhancement of the vacuum energy magnitude as $\Lambda$ increases. The right panel displays the 3D overlapping energy surfaces, providing a global perspective of the energy decay and the isotropic amplification induced by the longitudinal LV component $u_3$.}
  \label{casimir_IV_ab}
\end{figure}

\begin{figure}[h!]
    \centering
    \includegraphics[width=0.85\linewidth]{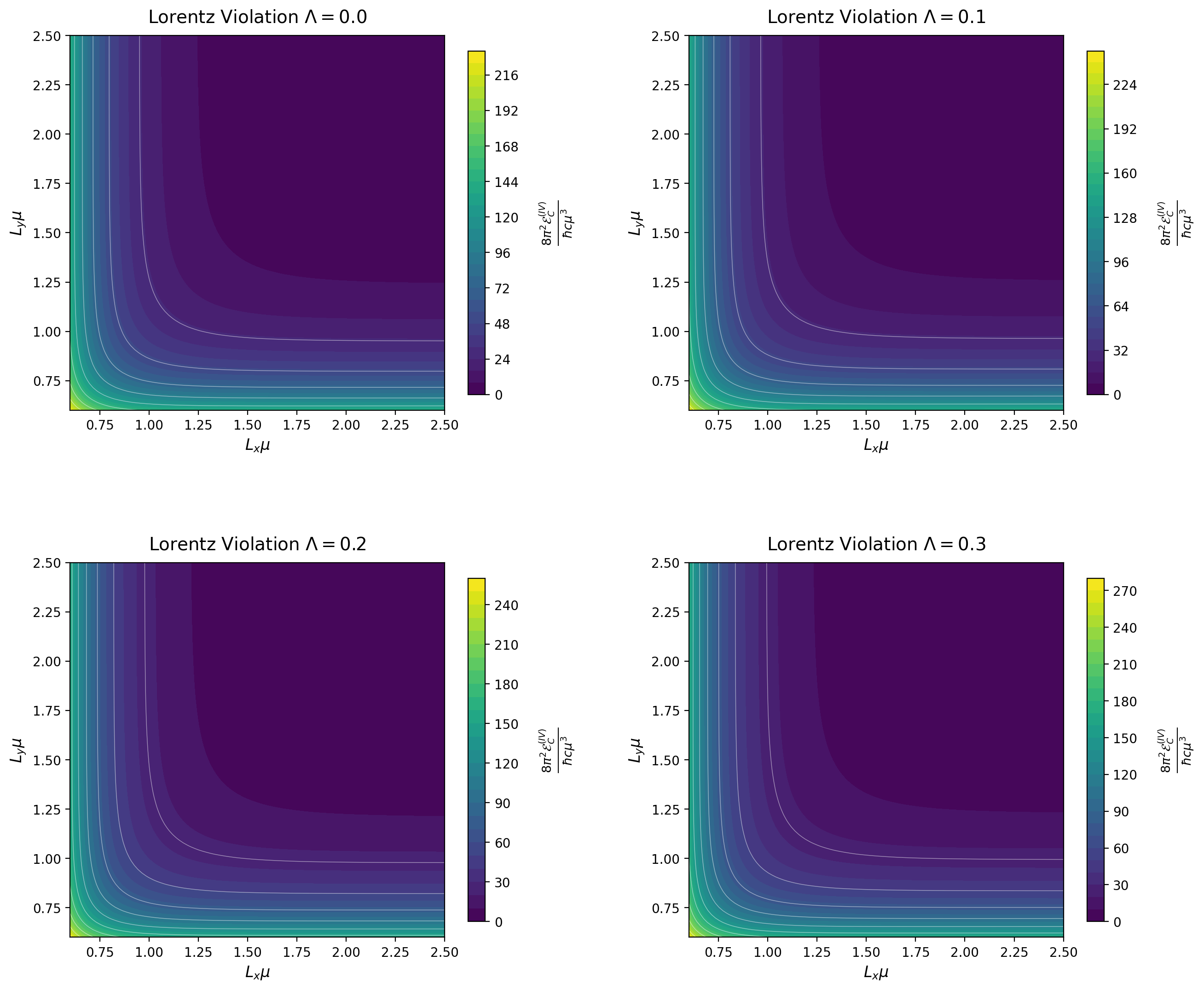}
    \caption{Contour maps of the Casimir energy density for Case IV, considering four distinct values of the LV parameter: $\Lambda = 0.0$ (standard Case), $0.1, 0.2,$ and $0.3$. The color gradient and the superimposed iso-energy lines represent the magnitude of $8 \pi^2 \mathcal{E}_C^{(IV)}/\hbar c \mu^3$. The perfect symmetry of the concentric contours across all panels demonstrates that Case IV preserves spatial isotropy in the $L_x$-$L_y$ plane, while the longitudinal LV component $u_3$ acts as a global amplification factor for the vacuum energy intensity.}
    \label{casimir_IV_c}
\end{figure}

The Casimir energy obtained from the Abel-Plana analysis does not admit a closed-form expression for generic values of the Lorentz-violating and field mass parameters. Therefore, in order to illustrate the physical consequences of Lorentz symmetry violation, we evaluate the resulting expressions numerically and present the Casimir energy as a function of the geometric parameters for the different Lorentz-violating configurations, we present the plots in Figs. \ref{casimir_I_ab}, \ref{casimir_I_c}, \ref{casimir_II_ab}, \ref{casimir_II_c}, \ref{casimir_III_ab}, \ref{casimir_III_c}, \ref{casimir_IV_ab} and \ref{casimir_IV_c}. In particular, we compare the isotropic limit with the timelike and spacelike background cases, highlighting how the anisotropic rescaling of the dispersion relation modifies the magnitude and scaling behavior of the vacuum energy. These numerical results provide a quantitative assessment of the impact of Lorentz violation on the Casimir effect and allow for a direct comparison between the different background orientations.

\subsection{Limiting cases} \label{limit_cases}

Now, in order to obtain analytical expressions, we consider the relevant asymptotic regimes of the theory, corresponding to the limits of large and small effective mass. To this end we sue the standard asymptotics of the modified Bessel function $K_{2}(z)$.

\begin{itemize}
    \item \textit{Small-mass limit:} For $z\ll1$ one has
\begin{equation}
K_{2}(z)
=
\frac{2}{z^{2}}
-\frac{1}{2}
+\mathcal{O}\!\big(z^{2}\ln z\big),
\qquad (z\to0^{+}).
\label{eq:K2_small}
\end{equation}
With $z=2p\mu\tilde L_x$, $z=2p\mu\tilde L_y$ and $z=2\mu R_{pq}$, this gives
\begin{equation}
\mu^{2}K_{2}(2\mu a)
=
\frac{1}{2a^{2}}
-\frac{\mu^{2}}{2}
+\mathcal{O}\!\big(\mu^{4}a^{2}\ln(\mu a)\big),
\qquad (a>0).
\label{eq:mu2K2_small}
\end{equation}
Substituting into \eqref{eq:Casimir_final_compact}, the leading (massless) part is
\begin{align}
\mathcal{E}^{(\lambda)}_{C}(L_x,L_y)
&=
\frac{\hbar c\,A^{(\lambda)}}{8\pi^2\sqrt{B^{(\lambda)}}}\,
\Bigg[
\frac{1}{2\tilde L_x^{3}}\sum_{p=1}^{\infty}\frac{1}{p^{4}}
+
\frac{1}{2\tilde L_y^{3}}\sum_{p=1}^{\infty}\frac{1}{p^{4}}
-
\frac{1}{2}\sum_{p,q\ge1}\frac{1}{R_{pq}^{4}}
\Bigg]
+\mathcal{O}\!\Big(\mu^{2}\ln\mu\Big)
\nonumber\\[4pt]
&=
\frac{\hbar c\,A^{(\lambda)}}{8\pi^2\sqrt{B^{(\lambda)}}}\,
\Bigg[
\frac{\pi^{4}}{180}\Big(\frac{1}{\tilde L_x^{3}}+\frac{1}{\tilde L_y^{3}}\Big)
-
\frac{1}{2}\sum_{p,q\ge1}\frac{1}{\big[(p\tilde L_x)^2+(q\tilde L_y)^2\big]^2}
\Bigg]
+\mathcal{O}\!\Big(\mu^{2}\ln\mu\Big),
\label{eq:EC_small_mu}
\end{align}
where we used $\sum_{p\ge1}p^{-4}=\pi^{4}/90$ and
$R_{pq}=\sqrt{(p\tilde L_x)^2+(q\tilde L_y)^2}$.
The correction $\mathcal{O}(\mu^{2}\ln\mu)$ originates from the
$\mathcal{O}(z^{2}\ln z)$ term in \eqref{eq:EC_small_mu}.

It is worth noting that the double series in Eq. \eqref{eq:EC_small_mu} is a particular case of a two-dimensional Epstein zeta function and can be reduced
to a rapidly convergent single-sum representation by performing one of the
integer sums analytically. In particular, using standard summation identities,
one obtains an equivalent form involving hyperbolic functions, which makes the
asymptotic behavior for large or small aspect ratio 
$\tilde L_y/\tilde L_x$ completely explicit. 
For instance, in the limit $\tilde L_y\gg\tilde L_x$ the leading behavior scales as
$\sim \zeta(3)/(\tilde L_x\,\tilde L_y^{3})$ (up to subleading algebraic and
exponentially small corrections), consistently with the expected approach
toward the parallel-plate configuration. 
We do not display the intermediate steps here, since this reduction follows
from standard properties of Epstein zeta functions.

\item \textit{Large-mass limit:} For $z\gg1$ one has
\begin{equation}
K_{2}(z)
\simeq
\sqrt{\frac{\pi}{2z}}\,e^{-z}\left(1+\frac{15}{8z}+\cdots\right),
\qquad (z\to+\infty).
\label{eq:K2_large}
\end{equation}
Therefore each series in \eqref{eq:Casimir_final_compact} is exponentially
suppressed, and the leading behavior is controlled by the smallest arguments,
namely $2\mu\tilde L_x$, $2\mu\tilde L_y$, and $2\mu R_{11}$:
\begin{align}
\mathcal{E}^{(\lambda)}_{C}(L_x,L_y)
&\simeq
\frac{\hbar c\,A^{(\lambda)}\,\mu^2}{8\pi^2\sqrt{B^{(\lambda)}}}
\Bigg[
\frac{1}{\tilde L_x}\,K_{2}(2\mu\tilde L_x)
+\frac{1}{\tilde L_y}\,K_{2}(2\mu\tilde L_y)
-\frac{1}{R_{11}^{2}}\,K_{2}(2\mu R_{11})
\Bigg]
\qquad (\mu\tilde L_{x,y}\gg1)
\nonumber\\[4pt]
&\simeq
\frac{\hbar c\,A^{(\lambda)}\,\mu^{3/2}}{8\pi^{3/2}\sqrt{B^{(\lambda)}}}
\Bigg[
\frac{e^{-2\mu\tilde L_x}}{\tilde L_x^{3/2}}
+
\frac{e^{-2\mu\tilde L_y}}{\tilde L_y^{3/2}}
-
\frac{e^{-2\mu R_{11}}}{R_{11}^{5/2}}
\Bigg]
\left[1+\mathcal{O}\!\Big(\frac{1}{\mu\,\min\{\tilde L_x,\tilde L_y\}}\Big)\right],
\label{eq:EC_large_mu}
\end{align}
with $R_{11}=\sqrt{\tilde L_x^{2}+\tilde L_y^{2}}$.
Thus the massive Casimir energy decays as $\sim e^{-2\mu\times(\text{distance})}$,
as expected for a gapped field.

\end{itemize}

\section{Conclusions and perspectives}
\label{conclusions}

In this work we investigated how Lorentz symmetry violation modifies the Casimir effect for a real massive scalar field confined to a rectangular waveguide with Dirichlet boundary conditions. The field dynamics was described by an aether-type, CPT-even deformation of the Klein-Gordon theory controlled by a small dimensionless parameter $\Lambda$ and a fixed background four-vector $u^\mu$. While the operator $(u \cdot \partial)^2$ generically induces mode mixing in the transverse basis, we focused on four representative aligned configurations of $u^\mu$ (one timelike and three spacelike) for which the spectral problem remains separable. In these cases LV enters entirely through anisotropic renormalizations of the temporal kinetic term (Case~I), the transverse confinement scales (Cases~II and~III), or the longitudinal propagation (Case~IV), encoded in the coefficients $A^{(\lambda)},B^{(\lambda)},C^{(\lambda)},D^{(\lambda)}$ of the unified dispersion relation \eqref{disp_general_case}.

Starting from the (formally divergent) vacuum energy density per unit length \eqref{E0_per_length_separable}, we implemented renormalization from the outset by isolating the geometry-dependent interaction energy through the twofold ``sum minus integral'' prescription \eqref{eq:EC_sum_minus_int_LV}. This subtraction removes the bulk contribution and the local boundary self-energies, so that the remaining quantity is finite and suitable for controlled analytic manipulations. We then evaluated the renormalized spectral bracket by a twofold application of the Abel-Plana formula, first over $n_x$ and subsequently over $n_y$. A key outcome of this procedure is that all potentially divergent local pieces generated by the Abel-Plana transformations are matched by the continuum subtraction in \eqref{eq:EC_sum_minus_int_LV}, while the physically relevant interaction energy is entirely governed by branch-cut discontinuities of the analytically continued spectrum. This leads to the compact finite representation \eqref{eq:EC_sum_minus_int_LV_fin}, in which the Casimir energy is written as a sum of manifestly nonlocal contributions that depend on the rescaled lengths $\tilde L_x=L_x/\sqrt{C^{(\lambda)}}$ and
$\tilde L_y=L_y/\sqrt{D^{(\lambda)}}$.

By exploiting the exponential series representation of the Bose-Einstein kernel and standard Bessel identities, we further reduced the result to an exponentially convergent Bessel-series form and performed analytically the remaining longitudinal momentum integral. The final Casimir energy density per unit length is given by the closed expression \eqref{eq:Casimir_final_compact}, which makes transparent how LV dresses the Lorentz-invariant answer: the overall normalization is rescaled by $A^{(\lambda)}/\sqrt{B^{(\lambda)}}$, while the transverse geometry enters through $\tilde L_x$ and $\tilde L_y$, i.e.\ through LV-dependent effective confinement lengths.

We also analyzed the limiting regimes of physical interest. In the small-mass limit, the energy reduces to a massless contribution expressed in terms of zeta values and a two-dimensional Epstein-zeta-type series, Eq.~\eqref{eq:EC_small_mu}, with the leading corrections controlled by the nonanalytic structure $\mathcal{O}(\mu^2\ln\mu)$. In the opposite large-mass regime, the energy is exponentially suppressed, Eq.~\eqref{eq:EC_large_mu}, with decay lengths governed by the LV-dressed scales $\tilde L_x$ and $\tilde L_y$. These asymptotics quantify how LV shifts the effective spectral gap and modifies the range and magnitude of vacuum stresses in confined geometries.

From a phenomenological viewpoint, the aligned configurations studied here provide a clean setting in which LV signatures can be isolated by comparing different orientations of $u^\mu$ and by monitoring how the Casimir energy and forces scale under controlled variations of $L_x$ and $L_y$. In particular, Cases~II and~III predict distinct responses under anisotropic rescalings of the transverse lengths, while Case~IV modifies the longitudinal dispersion and therefore the overall normalization through $B^{(\lambda)}$.

Several extensions follow naturally from the present framework. A first step is to treat generic orientations of $u^\mu$, where $(u \cdot \partial)^2$ induces genuine mode mixing and the spectral problem becomes matrix-valued in the transverse basis. It is also natural to consider alternative boundary conditions (e.g. Robin or mixed) and finite temperature, where Abel-Plana methods can be combined with Matsubara techniques to disentangle thermal corrections from LV-induced anisotropies. Finally, exploring other confined geometries and comparing with the parallel-plate limit would help identify which combinations of LV coefficients are most robustly encoded in observable Casimir forces.

\acknowledgements{A.M.-R. acknowledges financial support by UNAM-PAPIIT project No. IG100224, UNAM-PAPIME project No. PE109226, by SECIHTI project No. CBF-2025-I-1862 and by the Marcos Moshinsky Foundation.} E.R.B.M. thanks CNPq for partial support, Grant No. 304332/2024-0.

\appendix

\section{ Application of the Abel-Plana formula to the transverse mode sum in the $x$ direction } \label{appendix_AP_derivations}

In this Appendix we present the detailed evaluation of the transverse spectral contributions generated by the Abel-Plana transformation of the $x$-mode sum. We derive explicitly the local term, the continuum term, and the finite branch-cut remainder entering the renormalized transverse spectral sum defined in the main text.

Here we evaluate explicitly the contributions $\mathcal{S}^{(\lambda)}_{\mathrm{loc}} (k)$, $\mathcal{S}^{(\lambda)}_{\mathrm{cont}} (k)$ and $\mathcal{S}^{(\lambda)}_{\mathrm{AP}} (k)$ defined by Eqs.~\eqref{decomp_after_x_LV_new}, \eqref{Ex_cont_LV_new}, \eqref{ExAP_LV_new}, respectively.

By construction, the first term \eqref{decomp_after_x_LV_new} arises from the endpoint contribution $ - (1/2) F(0)$ of the Abel-Plana formula applied to the $n _{x}$ sum at fixed $n _{y}$ and $k$. The analytic continuation of the dispersion relation (\ref{disp_general_case}) yields
\begin{align}
\Omega ^{(\lambda)} (n _{y},k) \equiv \omega ^{(\lambda)} (0, n _{y},k) = c \, A ^{(\lambda)} \, \sqrt{ B ^{(\lambda)} \,  k ^{2} + D ^{(\lambda)} \, \left( \frac{n _{y} \pi}{L _{y}} \right) ^{2} + \left( \frac{mc}{\hbar} \right) ^{2} } ,  \label{disp_continuation}
\end{align}
and hence 
\begin{align}
    \mathcal{S}^{(\lambda)}_{\mathrm{loc}} (k) &= - \frac{1}{2} \sum _{n _{y} = 1} ^{\infty} \Omega ^{(\lambda)} (n _{y},k) . \label{Local_App} 
\end{align}
The second term \eqref{Ex_cont_LV_new} can be written as
\begin{align}
    \mathcal{S}^{(\lambda)}_{\mathrm{cont}} (k) &= \sum_{n_y=1}^{\infty} h ^{(\lambda)} (n _{y},k) ,
\end{align}
where
\begin{align}
    h ^{(\lambda)} (n _{y},k) = \int_0^{\infty}dn_x\; \omega ^{(\lambda)}(n_x,n_y;k) . \label{h_function}
\end{align}
To analyze the Abel-Plana remainder, we introduce the function
\begin{align}
    g ^{(\lambda)} (n _{y},k) = i
\int_0^{\infty}dt\;
\frac{
\omega^{(\lambda)}(it,n_y;k)
-
\omega^{(\lambda)}(-it,n_y;k)
}{e^{2\pi t}-1} , \label{g_function}
\end{align}
such that
\begin{align}
\mathcal{S}^{(\lambda)}_{\mathrm{AP}}(k)
&=
\sum_{n_y=1}^{\infty}  g ^{(\lambda)} (n _{y},k) . 
\label{SAP_start}
\end{align}
The function $g ^{(\lambda)} (n _{y},k)$ is entirely controlled by the analytic structure of the continued dispersion relation
$\omega^{(\lambda)}(z,n_y;k)$ as a function of the complex variable $z$.
For the separable spectrum considered here, given by Eq. \eqref{disp_general_case}, the analytic continuation $n_x\to z=\pm it$ produces the replacement
\begin{align}
\Big(\frac{\pi n_x}{L_x}\Big)^2
\ \longrightarrow\
\Big(\frac{\pi z}{L_x}\Big)^2
=
-\Big(\frac{\pi t}{L_x}\Big)^2,
\end{align}
which introduces a negative contribution that competes with the positive
mass-like terms proportional to
$B^{(\lambda)}k^2+D^{(\lambda)}(\pi n_y/L_y)^2+(mc/\hbar)^2$.
As a consequence, the argument of the square root in $\omega^{(\lambda)}(z,n_y;k)$
may change sign, and a branch cut is crossed only above a threshold value of $t$.

To make the analytic structure explicit, we introduce the $k$- and $n_{y}$-dependent gap function
\begin{align}
\Delta^{(\lambda)}(n_y,k;t)
\equiv
C^{(\lambda)}
\Big(\frac{\pi t}{L_x}\Big)^2
-
\Bigg[
B^{(\lambda)}k^2
+
D^{(\lambda)}\Big(\frac{\pi n_y}{L_y}\Big)^2
+
\Big(\frac{mc}{\hbar}\Big)^2
\Bigg].
\label{Delta_lambda_correct}
\end{align}
With this definition, the analytically continued dispersion relation becomes
\begin{align}
\omega^{(\lambda)}(\pm it,n_y;k)
=
c\,A^{(\lambda)}\,
\sqrt{
-
\Delta^{(\lambda)}(n_y,k;t)
}.
\label{omega_it_correct}
\end{align}
The threshold is determined by the condition
\begin{equation}
\Delta^{(\lambda)}(n_y,k;t^{(\lambda)} _{\ast})=0,
\end{equation}
which yields
\begin{align}
t^{(\lambda)} _{\ast} (n_y;k)
=
\frac{L_x}{\pi}
\sqrt{
\frac{
B^{(\lambda)}k^2
+
D^{(\lambda)}\big(\frac{\pi n_y}{L_y}\big)^2
+
\big(\frac{mc}{\hbar}\big)^2
}{C^{(\lambda)}}
}.
\label{t_threshold_correct}
\end{align}
For $0<t<t^{(\lambda)} _{\ast}(n_y;k)$ one has
$\Delta^{(\lambda)}(n_y,k;t)<0$,
so the argument of the square root in
\eqref{omega_it_correct} is positive and the continuation remains real. Hence $\omega^{(\lambda)}(it,n_y;k) = \omega^{(\lambda)}(-it,n_y;k)$, and the Abel-Plana integrand vanishes.

For $t>t^{(\lambda)} _{\ast} (n_y;k)$ one has
$\Delta^{(\lambda)}(n_y,k;t)>0$,
and the square root becomes purely imaginary.
Choosing the principal branch,
\begin{align}
\omega ^{(\lambda)} (it,n_y;k) &= + i c A ^{(\lambda)} \sqrt{\Delta ^{(\lambda)} (n_y,k;t)} , \nonumber \\ \omega ^{(\lambda)} (-it,n_y;k) &= -i c A ^{(\lambda)} \sqrt{\Delta^{(\lambda)} (n_y,k;t)}.
\end{align}
Therefore the discontinuity across the branch cut is
\begin{align}
\omega^{(\lambda)}(it,n_y;k) - \omega^{(\lambda)}(-it,n_y;k) = 2 i c A ^{(\lambda)} \sqrt{\Delta^{(\lambda)}(n_y,k;t)}, \qquad t > t ^{(\lambda)} _{\ast} (n_y,k). \label{disc_correct}
\end{align}
Substituting \eqref{disc_correct} into \eqref{SAP_start} yields the explicit cut form
\begin{align}
\mathcal{S}^{(\lambda)}_{\mathrm{AP}}(k) &= - 2 c A ^{(\lambda)} \sum_{n_y=1}^{\infty} \int_{t^{(\lambda) }_{\ast}(n_y,k)}^{\infty}dt\; \frac{ \sqrt{ C ^{(\lambda)}\big(\frac{\pi t}{L_x}\big)^2 -
\Big[
B^{(\lambda)}k^2
+
D^{(\lambda)}\big(\frac{\pi n_y}{L_y}\big)^2
+
\big(\frac{mc}{\hbar}\big)^2
\Big]
}
}{e^{2\pi t}-1}.
\label{SAP_cut_final}
\end{align}
Equation \eqref{SAP_cut_final} shows explicitly that the Abel-Plana remainder is finite: only the branch-cut discontinuity contributes, and the threshold $t ^{(\lambda)} _{\ast} (n_y,k)$ ensures that the square root is real in the integration domain.

\section{Second Abel-Plana transformation: summation over the $n_y$ modes} \label{appendix_AP_derivations_B}

In this Appendix we present the detailed analytic evaluation of the transverse spectral contributions associated with the discrete sum over the $y$-mode index $n_y$. This completes the twofold Abel-Plana analysis, following the evaluation of the $n_x$ sum presented in Appendix~\ref{appendix_AP_derivations}. In particular, we derive explicit expressions for the local term, the continuum contribution, and the finite branch-cut remainder appearing in the spectral decomposition introduced in the main text. Throughout this Appendix, the longitudinal momentum $k$ is treated as a fixed parameter.

\subsection{Local term}
\label{appendix_local_term}

We begin with the evaluation of the local contribution $\mathcal{S}^{(\lambda)}_{\mathrm{loc}} (k)$ defined in Eq.~\eqref{Local_App}. This term originates from the endpoint contribution of the Abel-Plana transformation of the $n_x$ sum and is therefore proportional to the residual discrete spectrum in the transverse $y$ direction.

The analytic continuation of the dispersion relation, $\Omega ^{(\lambda)} (n _{y},k)$, is given by Eq. \eqref{disp_continuation}. It is convenient to factor out the overall prefactor and define the $k$-dependent effective mass
\begin{align}
\Lambda ^{(\lambda)} (k) \equiv \sqrt{ B ^{(\lambda)} k ^{2} + \Big(\frac{mc}{\hbar}\Big) ^{2} } ,
\end{align}
so that
\begin{align}
\Omega ^{(\lambda)} (n _{y},k) = c \, A ^{(\lambda)} \sqrt{ [ \Lambda ^{(\lambda)} (k) ] ^{2} + D ^{(\lambda)} \Big(\frac{\pi n_y}{L_y}\Big) ^{2} } .
\end{align}
The local contribution can then be written as
\begin{align}
\mathcal{S} ^{(\lambda)}_{\mathrm{loc}}(k)
=
-\frac{cA^{(\lambda)}}{2}
\sum_{n_y=1}^{\infty}
\sqrt{
[ \Lambda ^{(\lambda)} (k) ] ^{2}
+
D^{(\lambda)}
\Big(\frac{\pi n_y}{L_y}\Big)^2
}.
\label{S_loc_start_appendix}
\end{align}
To evaluate the sum, we apply the Abel-Plana formula. Define the analytic continuation
\begin{align}
\xi ^{(\lambda)}(z;k) = \sqrt{ [ \Lambda ^{(\lambda)} (k) ] ^{2}  + D ^{(\lambda)} \Big(\frac{\pi z}{L _{y}} \Big) ^{2} } .
\end{align}
The Abel-Plana formula gives
\begin{align}
\sum_{n_y=1}^{\infty} \xi ^{(\lambda)}(n_y;k)
&=
-\frac12 \xi^{(\lambda)}(0;k)
+\int_0^\infty dn_y\,\xi^{(\lambda)}(n_y;k)
+i\int_0^\infty dt\;
\frac{
\xi^{(\lambda)}(it;k)-\xi^{(\lambda)}(-it;k)
}{e^{2\pi t}-1}.
\end{align}
The first two terms are straightforward:
\begin{align}
\xi ^{(\lambda)}(0;k)= \Lambda ^{(\lambda)}(k),
\qquad
\int_0^\infty dn_y\, \xi ^{(\lambda)}(n_y;k)
=
\frac{L_y}{\pi\sqrt{D^{(\lambda)}}}
\int_0^\infty d\chi\,
\sqrt{ [ \Lambda ^{(\lambda)} (k) ] ^{2} +\chi^2}.
\end{align}
For imaginary argument,
\begin{align}
\xi ^{(\lambda)}(it;k)
=
\sqrt{
[ \Lambda ^{(\lambda)} (k) ] ^{2}
-
D^{(\lambda)}
\Big(\frac{\pi t}{L_y}\Big)^2
},
\end{align}
which develops a branch cut when the argument of the square root
changes sign. This occurs for
\begin{align}
t>t^{(\lambda)}_{\ast \ast}(k)
=
\frac{L_y}{\pi}
\frac{\Lambda ^{(\lambda)}(k)}{\sqrt{D^{(\lambda)}}}.
\end{align}
Across the cut one finds
\begin{align}
\xi ^{(\lambda)}(it;k) - \xi ^{(\lambda)}(-it;k)
=
2i
\sqrt{
D^{(\lambda)}
\Big(\frac{\pi t}{L_y}\Big)^2
-
[ \Lambda ^{(\lambda)} (k) ] ^{2}
},
\qquad t>t^{(\lambda)}_{\ast \ast} (k).
\end{align}
Therefore,
\begin{align}
\sum_{n_y=1}^{\infty} \xi ^{(\lambda)}(n_y;k)
&=
-\frac{\Lambda ^{(\lambda)}(k)}{2}
+
\frac{L_y}{\pi\sqrt{D^{(\lambda)}}}
\int_0^\infty d\chi\,
\sqrt{ [ \Lambda ^{(\lambda)} (k) ] ^{2} + \chi^2}
-2\int_{t^{(\lambda)} _{\ast \ast}  (k)}^\infty dt\;
\frac{
\sqrt{
D^{(\lambda)}
\big(\frac{\pi t}{L_y}\big)^2
-
[ \Lambda ^{(\lambda)} (k) ] ^{2}
}
}{e^{2\pi t}-1}.
\end{align}
Substituting into Eq.~\eqref{S_loc_start_appendix}, the local spectral
contribution becomes
\begin{align}
\mathcal{S}^{(\lambda)}_{\mathrm{loc}}(k)
=
-\frac{cA^{(\lambda)}}{2}
\Bigg[
-\frac{   \Lambda ^{(\lambda)} (k)  }{2}
+
\frac{L_y}{\pi\sqrt{D^{(\lambda)}}}
\int_0^\infty d\chi\,
\sqrt{ [ \Lambda ^{(\lambda)} (k) ] ^{2} + \chi^2}
-
2\int_{t^{(\lambda)} _{\ast \ast} (k)}^\infty dt\;
\frac{
\sqrt{
D^{(\lambda)}
\big(\frac{\pi t}{L_y}\big)^2
-
[ \Lambda ^{(\lambda)} (k) ] ^{2}
}
}{e^{2\pi t}-1}
\Bigg].
\end{align}
This expression separates the local spectral contribution into an endpoint term, a continuum integral, and a finite branch-cut contribution generated by the Abel-Plana remainder.

\subsection{Continuum term}
\label{appendix_cont_term}

We now evaluate the continuum contribution
\begin{align}
\mathcal{S}^{(\lambda)}_{\mathrm{cont}}(k)
=
\sum_{n_y=1}^{\infty} h^{(\lambda)}(n_y,k),
\end{align}
with
\begin{align}
h^{(\lambda)}(n_y,k)
=
\int_0^\infty dn_x\;
\omega^{(\lambda)}(n_x,n_y;k),
\end{align}
where the dispersion relation is
\begin{align}
\omega^{(\lambda)}(n_x,n_y;k)
=
cA^{(\lambda)}
\sqrt{
B^{(\lambda)}k^2
+
C^{(\lambda)}\Big(\frac{\pi n_x}{L_x}\Big)^2
+
D^{(\lambda)}\Big(\frac{\pi n_y}{L_y}\Big)^2
+
\Big(\frac{mc}{\hbar}\Big)^2 } .
\end{align}
Factoring out the overall prefactor and introducing
\begin{align}
\Lambda^{(\lambda)}(k)
\equiv
\sqrt{
B^{(\lambda)}k^2
+
\Big(\frac{mc}{\hbar}\Big)^2
},
\end{align}
we define the analytic continuation
\begin{align}
h^{(\lambda)}(z,k)
=
cA^{(\lambda)}
\int_0^\infty dn_x\;
\sqrt{
[ \Lambda ^{(\lambda)} (k) ] ^{2} 
+
C^{(\lambda)}\Big(\frac{\pi n_x}{L_x}\Big)^2
+
D^{(\lambda)}\Big(\frac{\pi z}{L_y}\Big)^2 } .
\end{align}
Applying the Abel-Plana formula to the sum over $n_y$ gives
\begin{align}
\sum_{n_y=1}^{\infty} h^{(\lambda)}(n_y,k)
&=
-\frac12 h^{(\lambda)}(0,k)
+\int_0^\infty dn_y\,h^{(\lambda)}(n_y,k)
+i\int_0^\infty dt\;
\frac{
h^{(\lambda)}(it,k)-h^{(\lambda)}(-it,k)
}{e^{2\pi t}-1}.
\label{AP_cont_start}
\end{align}
The first two terms are
\begin{align}
h^{(\lambda)}(0,k)
&= c A^{(\lambda)} \frac{L_x}{\pi\sqrt{C^{(\lambda)}}}
\int_0^\infty d\chi\,
\sqrt{ [ \Lambda ^{(\lambda)} (k) ] ^{2} + \chi^2} ,
\\[5pt]
\int_0^\infty dn_y\,h^{(\lambda)}(n_y,k)
&=
\int_0^\infty dn_y
\int_0^\infty dn_x \;
\omega^{(\lambda)}(n_x,n_y;k).
\end{align}
To evaluate the Abel-Plana remainder, set $z=it$. Then
\begin{align}
\omega^{(\lambda)}(n_x,it;k)
=
cA^{(\lambda)}
\sqrt{
[ \Lambda ^{(\lambda)} (k) ] ^{2}
+
C^{(\lambda)}\Big(\frac{\pi n_x}{L_x}\Big)^2
-
D^{(\lambda)}\Big(\frac{\pi t}{L_y}\Big)^2 } .
\end{align}
We now introduce the $t$-dependent quantity
\begin{align}
\Xi^{(\lambda)}(t,k)
=
D^{(\lambda)}\Big(\frac{\pi t}{L_y}\Big)^2
-
[ \Lambda ^{(\lambda)} (k) ] ^{2} .
\end{align}
A branch cut appears when $\Xi^{(\lambda)}(t,k)>0$, i.e.
\begin{align}
t>t^{(\lambda)}_{\ast\ast}(k)
=
\frac{L_y}{\pi}
\frac{\Lambda^{(\lambda)}(k)}{\sqrt{D^{(\lambda)}}}.
\end{align}
For fixed $t>t^{(\lambda)}_{\ast\ast}(k)$ the square root becomes
imaginary for
\begin{align}
0<n_x<n_{\ast}(t,k),
\qquad
n_{\ast}(t,k)
=
\frac{L_x}{\pi}
\sqrt{\frac{\Xi^{(\lambda)}(t,k)}{C^{(\lambda)}}}.
\end{align}
Across the cut,
\begin{align}
\omega^{(\lambda)}(n_x,it;k)
&=
+i cA^{(\lambda)}
\sqrt{
\Xi^{(\lambda)}(t,k)
-
C^{(\lambda)}\Big(\frac{\pi n_x}{L_x}\Big)^2},
\\
\omega^{(\lambda)}(n_x,-it;k)
&=
-i cA^{(\lambda)}
\sqrt{
\Xi^{(\lambda)}(t,k)
-
C^{(\lambda)}\Big(\frac{\pi n_x}{L_x}\Big)^2}.
\end{align}
Therefore
\begin{align}
h^{(\lambda)}(it,k)-h^{(\lambda)}(-it,k)
&=
2i cA^{(\lambda)}
\int_0^{n_{\ast}(t,k)} dn_x\;
\sqrt{
\Xi^{(\lambda)}(t,k)
-
C^{(\lambda)}\Big(\frac{\pi n_x}{L_x}\Big)^2}.
\end{align}
Using the standard semicircle integral,
\begin{align}
\int_0^{n_{\ast}} dn_x\;
\sqrt{
\Xi
-
C^{(\lambda)}\Big(\frac{\pi n_x}{L_x}\Big)^2}
=
\frac{L_x}{4\sqrt{C^{(\lambda)}}}\,\Xi,
\end{align}
we obtain
\begin{align}
h^{(\lambda)}(it,k)-h^{(\lambda)}(-it,k)
=
i\,cA^{(\lambda)}
\frac{L_x}{2\sqrt{C^{(\lambda)}}}
\Xi^{(\lambda)}(t,k),
\qquad t>t^{(\lambda)}_{\ast\ast}(k),
\end{align}
and zero otherwise.

Substituting into the Abel-Plana remainder gives
\begin{align}
i\int_0^\infty dt\;
\frac{
h^{(\lambda)}(it,k)-h^{(\lambda)}(-it,k)
}{e^{2\pi t}-1}
=
-\frac{cA^{(\lambda)}L_x}{2\sqrt{C^{(\lambda)}}}
\int_{t^{(\lambda)}_{\ast\ast}(k)}^\infty dt
\frac{
D^{(\lambda)}\Big(\frac{\pi t}{L_y}\Big)^2
-
[ \Lambda ^{(\lambda)} (k) ] ^{2}
}{e^{2\pi t}-1}.
\end{align}
The continuum spectral contribution is therefore
\begin{align}
\mathcal{S}^{(\lambda)}_{\mathrm{cont}}(k)
=
- \frac{c A^{(\lambda)}}{2}  \left[ \frac{L_x}{\pi\sqrt{C^{(\lambda)}}}
\int_0^\infty d\chi\,
\sqrt{ [ \Lambda ^{(\lambda)} (k) ] ^{2} + \chi^2}
+ \frac{ L_x}{ \sqrt{C^{(\lambda)}}}
\int_{t^{(\lambda)}_{\ast\ast}(k)}^\infty dt
\frac{
D^{(\lambda)}\Big(\frac{\pi t}{L_y}\Big)^2
-
[ \Lambda ^{(\lambda)} (k) ] ^{2}
}{e^{2\pi t}-1} \right] \notag \\[5pt] + \int_0^\infty dn_y
\int_0^\infty dn_x \;
\omega^{(\lambda)}(n_x,n_y;k)  .
\end{align}

\subsection{Second Abel-Plana transformation of the branch-cut contribution} \label{appendix_second_AP}

We now evaluate the remaining spectral contribution arising from the Abel-Plana remainder of the first transverse summation. This term still contains a discrete sum over the transverse index $n_y$, which must be treated by applying the Abel-Plana formula once more.

From the result of the first summation, the remaining contribution can be written as
\begin{align}
\mathcal{S}^{(\lambda)}_{\mathrm{AP}} (k) &=  \sum_{n_y=1}^{\infty} g^{(\lambda)}(n_y,k),
\end{align}
where
\begin{align}
g^{(\lambda)}(z,k)
= - 2 c A ^{(\lambda)}
\int_{t^{(\lambda)}_{\ast}(z,k)}^\infty dt\;
\frac{
\sqrt{
C^{(\lambda)}\Big(\frac{\pi t}{L_x}\Big)^2
-
[ \Lambda ^{(\lambda)} (k) ] ^{2}
-
D^{(\lambda)}\Big(\frac{\pi z}{L_y}\Big)^2
}
}{e^{2\pi t}-1},
\end{align}
and the lower integration limit is determined by the threshold condition
\begin{align}
t^{(\lambda)}_{\ast}(z,k)
=
\frac{L_x}{\pi}
\sqrt{
\frac{
[ \Lambda ^{(\lambda)} (k) ] ^{2}
+
D^{(\lambda)}\Big(\frac{\pi z}{L_y}\Big)^2
}{C^{(\lambda)}}
}.
\end{align}
Applying the Abel-Plana formula to the sum over $n_y$ gives
\begin{align}
\sum_{n_y=1}^{\infty} g^{(\lambda)}(n_y,k)
&=
 g^{(\lambda)}(0,k)
+\int_0^\infty dn_y\,g^{(\lambda)}(n_y,k)
+i\int_0^\infty dt\;
\frac{
g^{(\lambda)}(it,k)-g^{(\lambda)}(-it,k)
}{e^{2\pi t}-1}.
\label{AP_second_start}
\end{align}
The first term follows directly from the definition of $g^{(\lambda)}$,
\begin{align}
-\frac{1}{2} g^{(\lambda)}(0,k)
=
  c A ^{(\lambda)}
\int_{t^{(\lambda)}_{\ast}(0,k)}^\infty dt\;
\frac{
\sqrt{
C^{(\lambda)}\Big(\frac{\pi t}{L_x}\Big)^2
-
[ \Lambda ^{(\lambda)} (k) ] ^{2}
}
}{e^{2\pi t}-1}.
\end{align}
The second term is evaluated by rescaling the transverse variable
$\chi=\pi n_y/L_y$ and interchanging the order of integration, which yields
\begin{align}
\int_0^\infty dn_y\,g^{(\lambda)}(n_y,k)
= -  
\frac{ c A ^{(\lambda)} L_y}{2\sqrt{D^{(\lambda)}}}
\int_{t^{(\lambda)}_{\ast}(0,k)}^\infty dt\;
\frac{
C^{(\lambda)}\Big(\frac{\pi t}{L_x}\Big)^2
-
[ \Lambda ^{(\lambda)} (k) ] ^{2}
}{e^{2\pi t}-1}.
\end{align}
To evaluate the third term, we analyze the analytic structure of
$g^{(\lambda)}(z,k)$ under the continuation $z\to\pm it$. Introducing the
variable $u=\pi\tau/L_x$ and defining
\begin{align}
M^{(\lambda)}(z,k)
=
\sqrt{
[ \Lambda ^{(\lambda)} (k) ] ^{2}
+
D^{(\lambda)}\Big(\frac{\pi z}{L_y}\Big)^2},
\end{align}
the function $g^{(\lambda)}$ can be written as
\begin{align}
g^{(\lambda)}(z,k)
=
\frac{L_x}{\pi\sqrt{C^{(\lambda)}}}
\int_{M^{(\lambda)}(z,k)}^\infty du\;
\frac{\sqrt{u^2- [M^{(\lambda)}(z,k)]^2}}
{e^{2L_x u/\sqrt{C^{(\lambda)}}}-1}.
\end{align}
For imaginary argument $z=\pm it$ one obtains
\begin{align}
M^{(\lambda)}(\pm it,k)
=
\sqrt{
[ \Lambda ^{(\lambda)} (k) ] ^{2}
-
D^{(\lambda)}\Big(\frac{\pi t}{L_y}\Big)^2}.
\end{align}
The square root becomes imaginary when
\begin{align}
t>t^{(\lambda)}_{\ast\ast}(k)
=
\frac{L_y}{\pi}
\frac{\Lambda^{(\lambda)}(k)}{\sqrt{D^{(\lambda)}}}.
\end{align}
For $t$ below this threshold the function is analytic and the Abel-Plana integrand vanishes. For $t$ above threshold we define
\begin{align}
B^{(\lambda)}(t,k)
=
\sqrt{
D^{(\lambda)}\Big(\frac{\pi t}{L_y}\Big)^2
-
[ \Lambda ^{(\lambda)} (k) ] ^{2} },
\end{align}
and crossing the principal branch cut gives
\begin{align}
M^{(\lambda)}(it,k)=+iB^{(\lambda)}(t,k),
\qquad
M^{(\lambda)}(-it,k)=-iB^{(\lambda)}(t,k).
\end{align}
The discontinuity is obtained from the integral along the branch cut in the $u$ plane. Parametrizing the cut by $u=is$ with $s\in[-B^{(\lambda)}(t,k),B^{(\lambda)}(t,k)]$ yields
\begin{align}
g^{(\lambda)}(it,k)-g^{(\lambda)}(-it,k)
=
\frac{2iL_x}{\pi\sqrt{C^{(\lambda)}}}
\int_{-B^{(\lambda)}(t,k)}^{B^{(\lambda)}(t,k)} ds\;
\sqrt{ [B^{(\lambda)}(t,k)] ^2-s^2}
\frac{1}{e^{2iL_x s/\sqrt{C^{(\lambda)}}}-1},
\qquad t>t^{(\lambda)}_{\ast\ast}(k).
\end{align}
The Abel-Plana remainder therefore reduces to
\begin{align}
i\int_0^\infty dt\;
\frac{g^{(\lambda)}(it,k)-g^{(\lambda)}(-it,k)}{e^{2\pi t}-1}
=
i\int_{t^{(\lambda)}_{\ast\ast}(k)}^\infty dt\;
\frac{g^{(\lambda)}(it,k)-g^{(\lambda)}(-it,k)}{e^{2\pi t}-1}.
\end{align}
Collecting all contributions, the second Abel-Plana transformation gives
\begin{align}
\sum_{n_y=1}^{\infty} g^{(\lambda)}(n_y,k)
&=
-\frac{1}{2}
\int_{t^{(\lambda)}_{\ast}(0,k)}^\infty dt\;
\frac{
\sqrt{
C^{(\lambda)}\Big(\frac{\pi t}{L_x}\Big)^2
-
[ \Lambda ^{(\lambda)} (k) ] ^{2}
}
}{e^{2\pi t}-1}
+\frac{L_y}{4\sqrt{D^{(\lambda)}}}
\int_{t^{(\lambda)}_{\ast}(0,k)}^\infty dt\;
\frac{
C^{(\lambda)}\Big(\frac{\pi t}{L_x}\Big)^2
-
[ \Lambda ^{(\lambda)} (k) ] ^{2}
}{e^{2\pi t}-1} \notag \\ & \hspace{3cm}
+i\int_{t^{(\lambda)}_{\ast\ast}(k)}^\infty dt\;
\frac{g^{(\lambda)}(it,k)-g^{(\lambda)}(-it,k)}{e^{2\pi t}-1}.
\end{align}


\bibliography{references}

\end{document}